\documentclass[12pt, a4paper]{article}
  \usepackage{algpseudocode}
  \usepackage{algorithm}
  \usepackage{amsfonts}
  \usepackage{amsmath}
  \usepackage{amssymb}
  \usepackage{amsthm}
  \usepackage{array}
  \usepackage{bm}
  \usepackage{booktabs}
  \usepackage{caption}
  \usepackage{color}
  \usepackage{float}
  \usepackage[stable, bottom, marginal]{footmisc}
  \usepackage{geometry}
  \usepackage{graphicx}
  \usepackage{algorithm}
  \usepackage{lineno}
  \usepackage{mathrsfs}
  \usepackage{multirow}
  \usepackage{natbib}
  \usepackage[normalem]{ulem}
  \usepackage{subfigure}
  \usepackage{url}
  \usepackage{ulem}
  \usepackage{epstopdf}
  \usepackage{bm}
  \usepackage{tabularx}
  \usepackage{setspace}
  \usepackage{appendix}
  \usepackage{verbatim}
  \setcitestyle{authoryear, open = {(}, close = {)}}
  \theoremstyle{plain}
  \newtheorem{definition}{Definition}
  \newtheorem{proposition}{Proposition}
  
  \theoremstyle{definition}
  
\begin{document}

\begin{center}
  {\bfseries \Large
       Dimension reduction of open-high-low-close data in candlestick chart based on pseudo-PCA
  }
\end{center}

\vskip 0.3cm
\begin{center}
Wenyang Huang$^{1, 2}$, Huiwen Wang$^{1, 3}$ and Shanshan Wang$^{1, 3*}$ \\
\vskip 0.3cm
\small
$^1$\emph{School of Economics and Management, Beihang University, Beijing 100191, China} \\
$^2$\emph{Shen Yuan Honors College, Beihang University, Beijing 100191, China} \\
$^3$\emph{Beijing Advanced Innovation Center for Big Data and Brain Computing, Beijing 100191, China}
\end{center}

\let \thefootnote \relax \footnotetext{
* Corresponding author. Correspondence to: School of Economics and Management, Beihang University, Beijing 100191, China.
E-mail address: sswang@buaa.edu.cn (S.S. Wang).}

\vskip 0.3cm
\begin{center}
  {\bfseries \large
    Abstract}
\end{center}
\vspace*{-0.3cm}

The (open-high-low-close) OHLC data is the most common data form in the field of finance and the investigate object of various technical analysis.
With increasing features of OHLC data being collected, the issue of extracting their useful information in a comprehensible way for visualization and easy interpretation must be resolved.
The inherent constraints of OHLC data also pose a challenge for this issue.
This paper proposes a novel approach to characterize the features of OHLC data in a dataset and then performs dimension reduction, which integrates the feature information extraction method and principal component analysis.
We refer to it as the pseudo-PCA method.
Specifically, we first propose a new way to represent the OHLC data, which will free the inherent constraints and provide convenience for further analysis.
Moreover, there is a one-to-one match between the original OHLC data and its feature-based representations, which means that the analysis of the feature-based data can be reversed to the original OHLC data.
Next, we develop the pseudo-PCA procedure for OHLC data, which can effectively identify important information and perform dimension reduction. Finally, the effectiveness and interpretability of the proposed method are investigated through finite simulations and the spot data of China's agricultural product market.

\vskip 0.5cm
\noindent
  {\bfseries
    Key Words:}
    OHLC data;
    pseudo-PCA;
    dimension reduction;
    feature information extraction

\section{Introduction} \label{Sec 1}

As one of the most classic technical analysis tools, the candlestick chart is used to document the prices of almost all financial products, including spots, stocks, futures, options, etc \citep{romeo2015study, chmielewski2015pattern}. Investors can judge the long and short trading conditions and roughly predict future price trends by analyzing candlestick chart \citep{Tsai2014Stock}.

\begin{figure}[!h]
  \centering
  \resizebox{17cm}{7cm}{\includegraphics{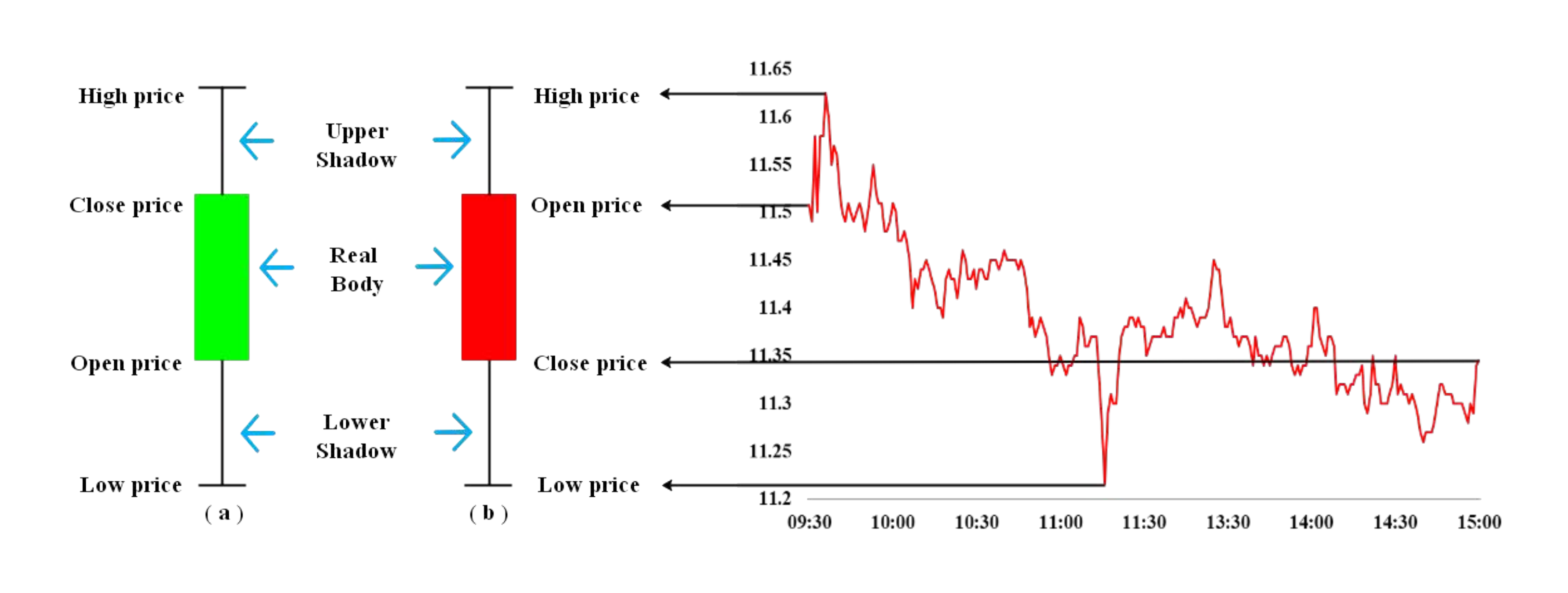}}
  \caption{An example of daily candlestick chart}\label{Fig candlestickchart}
\end{figure}

Intuitively, Figure.\ref{Fig candlestickchart} shows a daily candlestick chart, which not only records the open price, high price, low price, and close price of certain object on that day, but also reflects the difference between any two prices.
According to American convention, a green real body refers to a bull market (the close price $\textgreater$ the open price, see Figure.\ref{Fig candlestickchart}(a)), while a red real body refers to a bear market (the open price $\textgreater$ the close price, see Figure.\ref{Fig candlestickchart}(b)), respectively.
\vspace*{-0.6cm}
\vspace*{-0.5cm}

Mathematically, the essence of the candlestick chart is a $4$-dimensional vector, consisting of
the \underline{o}pen price, \underline{h}igh price, \underline{l}ow price, and \underline{c}lose price, which are collectively called the OHLC data \citep{kazemilari2015correlation}.
However, unlike ordinary $4$-dimensional vectors, the OHLC data have some important inherent constraints. Specifically,
\begin{definition}\label{OHLC data}
A $4$-dimensional vector $\bm{x}=(x^{(o)}, x^{(h)}, x^{(l)}, x^{(c)})'$ is considered the OHLC data if $\bm{x}$ satisfies the following three constraints:
\begin{center}
  (1) $x^{(l)} > 0$;\\
  (2) $x^{(l)} < x^{(h)}$; \\
  (3) $x^{(o)}, x^{(c)} \in [x^{(l)},x^{(h)}]$.
\end{center}
where $x^{(o)}, x^{(h)},x^{(l)}$ and $x^{(c)}$ represent the open price,  high price, low price, and close price of the subject during a certain time period (which can vary from seconds to months), respectively.
\end{definition}
\vspace*{-0.2cm}

The OHLC data $\bm{x}$ in Definition.\ref{OHLC data} can be defined as a generalized interval-valued data from the perspective of symbolic data analysis \citep{article}. Namely, the OHLC data takes $x^{(h)}$ and $x^{(l)}$ as the upper and lower boundaries of the interval, while $x^{(o)}$ and $x^{(c)}$ lie within the interval.
Compared to the interval-valued data, the OHLC data contains more information, resulting in more complicated properties than interval-valued data. The common methods for interval-valued data, such as the Max $\&$ Min method \citep{arroyo2011different} and Center $\&$ Range \citep{brito2012modelling, giordani2015lasso} method are mimicked by various works on the OHLC data. For example, \cite{yang2012acix}, \cite{xiong2017interval}, and \cite{sun2018threshold} have investigated the OHLC data via the above methods used for interval-valued data without incorporating information on the open price and close price.

Owing to the ignorance of the open price and close price, these two widely-used methods cannot completely use the OHLC data information, yielding possible inefficient results. As indicated by \cite{cheung2007empirical}, the open price and close price have an explanatory power for the fluctuation of the OHLC data and should be carefully incorporated into statistical modeling.

Through decades of development, scholars have completed multi-angle studies on the candlestick chart and its data  (i.e., OHLC data). Initially, scholars are committed to performing pattern recognition on the candlestick chart \citep{kamijo1990stock, juang1992discriminative, lo2000foundations, cervello2015stock}, giving priority to finding a candlestick chart pattern that could obtain stable profits. At the same time, traditional statistical methods such as the clustering analysis \citep{harris1991stock, brown2002influence, tao2017k} and discriminant analysis \citep{hopkins1996effect, leung2000forecasting, cohen2019trading} have also been gradually applied to the study of the candlestick chart.
In addition, other popular researches include predicting the future changes of OHLC data, such as econometric forecasting models \citep{fiess2002towards, pai2005hybrid, de2020large} and machine learning models \citep{cao2003support, liu2012fluctuation, mann2017new, mallqui2019predicting}.

With technological advances in many scientific areas, a considerable amount of OHLC data is available. It is crucial to develop a dimension reduction technique to extract the informative features contained in the multivariate OHLC data.  Among the dimension reduction methods, the principal component analysis (PCA) has always been a key technique to extract the comprehensive features using the linear combination of the original variables, which are called principal components (PCs). These PCs represent the major mode of data variation and are the characterizing features of typical variables in a dataset \citep{wang2015principal,m2016forecasting}. Furthermore, the scores on PCs have been used for visualization to help researchers better understand the correlation between the original variables \citep{jolliffe2002principal}.

Nonetheless, the aforementioned PCA method is not applicable when handling dimension reduction for the OHLC data due to its inherent constraints, which poses new challenges. For example, imposing the classical PCA on the OHLC dataset will destroy its structure, rendering the corresponding PCs meaningless.  Furthermore, as a special type of data, the OHLC data is characterized as a $4$-dimensional vector data with constraints defined in Definition.\ref{OHLC data} as a whole, which is completely different from the classical scalar data or the vector data in the Euclidean space. All of these concerns are motivations to propose a new approach.

As a reminder, the OHLC data can be used as generalized interval-valued data. Fortunately, various PCA techniques have been extended from classical scalar data to the interval-valued data, such as the vertices principal component analysis (VPCA; \cite{cazes1997extension}) , centers principal component analysis (CPCA; \cite{chouakria2000symbolic}), symbolic object principal component analysis (SPCA; \cite{lauro2000principal}), midpoints radii principal component analysis (MRPCA; \cite{palumbo2003pca}), interval principal component analysis (IPCA; \cite{gioia2006principal}), and complete-information-based principal component analysis (CIPCA; \cite{wang2012cipca}).
After revisiting the aforementioned PCA methods for interval-valued data, we find that the key point is to extract feature-based representation of interval-valued data, such as vertices, midpoint, radii, etc.  Nonetheless, these existing approaches are only designed for interval-valued data, and they could not be directly applied to the OHLC data.
Apart from the two boundaries contained in the interval-valued data, the OHLC data includes more information and constraints.  Consequently, this requires more feature-based representations.

Therefore, this paper aims to explore dimension reduction for the OHLC data based on the PCA method with the idea of feature-based representations, constructing a novel approach to characterize features of OHLC data. First, we propose a new way to represent the OHLC data, which can eliminate the inherent constraints and make further analysis easier. Furthermore, there is a one-to-one match between the original OHLC data and its feature-based representations, which allows the analysis of the feature-based representations to be reversed to the original data. Next, we develop the pseudo-PCA procedure for OHLC data, which can effectively identify important information and perform dimension reduction. Finally, the effectiveness and interpretability of the proposed method are investigated through finite simulations and the spot data of China's agricultural product market.

This paper has two main contributions. First, a novel feature-based representation method is proposed for the OHLC data, which provides a new perspective and facilitates further statistical analysis of the OHLC data. Second, we introduce a pseudo-PCA procedure to achieve dimension reduction for multivariate OHLC data, which enhances existing literature on PCA methods.

The rest of the paper is organized as follows: Section \ref{Sec 2} presents the feature extraction method for the OHLC data and related algebraic space of the feature-based representations. In Section \ref{Sec 3}, we describe the proposed pseudo-PCA procedure.  Then we examine the effectiveness, robustness and interpretability of the pseudo-PCA by simulations in Section.\ref{Sec simulation} and empirical experiment in Section \ref{Sec 4}. Finally, we conclude with a brief summary and insights, as well as several possible further research topics in Section \ref{Sec 5}.

\vspace*{-0.8cm}
\section{Feature-based representations and its algebraic space} \label{Sec 2}
\vspace*{-0.5cm}

\subsection{\textit{Feature-based representations of OHLC data}} \label{Sec_2.1}

Inspired by the feature extraction method for interval-valued data, we propose a novel feature-based representation approach for the OHLC data in Definition.\ref{feature information}. This satisfies the following merits: First, each component of feature-based representations has a unique and meaningful interpretation of the OHLC data. Second, it eliminates the inherent constraints and provides convenience for further analysis. Third, there is a one-to-one match between the original OHLC data and its feature-based representations, which allows the analysis of the feature-based representations to be reversed to the original data.
\vspace*{-0.2cm}

\begin{definition}\label{feature information}
(\textbf{Feature-based representations of OHLC data}). For the OHLC data $\bm{x}=(x^{(o)}, x^{(h)}, x^{(l)}, x^{(c)})'$,
assume that $x^{(o)}, x^{(h)}, x^{(l)}$, and $x^{(c)}$ are not equal to each other $(\text{except for} \; x^{(o)} \equiv x^{(c)})$,  then the feature-based representation of $\bm{x}$ is formulated as $\bm{y}=(y^{(1)},y^{(2)},y^{(3)},y^{(4)})'$, where
\begin{equation}\label{Eq_1}
 \left\{\begin{array}{l}
  y^{(1)}=\ln x^{(l)} \\
  y^{(2)}=\ln (x^{(h)}-x^{(l)}) \\
  y^{(3)}=\ln\Big(\frac{\lambda^{(o)}}{1-\lambda^{(o)}}\Big) \\
  y^{(4)}=\ln\Big(\frac{\lambda^{(c)}}{1-\lambda^{(c)}}\Big) \\
\end{array}\right.,
\end{equation}
 with
\begin{equation}\label{Eq_2}
\lambda^{(o)} = \frac{x^{(o)}-x^{(l)}}{x^{(h)}-x^{(l)}} \ \ \ \mbox{and}\ \ \ \lambda^{(c)} = \frac{x^{(c)}-x^{(l)}}{x^{(h)}-x^{(l)}}.
\end{equation}
\end{definition}

From Definition.\ref{feature information}, each component of the feature-based representation for the OHLC $\bm{x}$ is significant. Specifically, $y^{(1)}$ corresponds to the natural logarithmic value of the low price, reflecting the magnitude of the original OHLC data.
$y^{(2)}$ is the natural logarithmic value of the range between the high price and the low price, which can note the fluctuation of raw OHLC data.
Obviously, $y^{(1)}$ and $y^{(2)}$ can take their values freely in $[-\infty, +\infty]$.
Furthermore, in order to ensure that the constraint $x^{(o)}, x^{(c)} \in [x^{(l)},x^{(h)}]$ always holds, we introduce two convex combination coefficients $\lambda^{(o)}$ and $\lambda^{(c)}$, which link $\lambda^{(o)}$ and $x^{(o)}$ as well as $\lambda^{(c)}$ and $x^{(c)}$ together with $x^{(l)}$ and $x^{(h)}$. That is,
\vspace*{-0.5cm}
\begin{equation}\label{Eq_lambda to x}
  x^{(o)}=\lambda^{(o)}x^{(h)}+(1-\lambda^{(o)})x^{(l)} \ \ \ \mbox{and}\ \ \
  x^{(c)}=\lambda^{(c)}x^{(h)}+(1-\lambda^{(c)})x^{(l)}.
\end{equation}
\vspace*{-1.5cm}

$\lambda^{(o)},\lambda^{(c)} \in (0,1)$ implies $x^{(o)}, x^{(c)} \in [x^{(l)},x^{(h)}]$.
Besides, the implication of $\lambda^{(o)}$ and $\lambda^{(c)}$ is very clear.
Specifically, when $\lambda^{(o)} \textgreater 0.5$, it implies that $x^{(o)}$ is closer to $x^{(h)}$;
and if $\lambda^{(o)} \textless 0.5$, it indicates that $x^{(o)}$ is closer to $x^{(l)}$.
The interpretation of $\lambda^{(c)}$ is similar. Note that $\lambda^{(o)}$ and $\lambda^{(c)}$ are limited to $(0,1)$, which mimic the proportion or probability of $x^{(o)}$ and $x^{(c)}$, respectively, which measures the relative distance from $x^{(l)}$. Thus, similar to the idea of logistic regression, we utilize the ``log odds ratio" to yield $y^{(3)}$ and $y^{(4)}$, which also removes the finite support $(0,1)$ constraint and free $y^{(3)}$ and $y^{(4)}$ to $[-\infty, +\infty]$.

Furthermore, a one-to-one correspondence between the original OHLC data $\bm{x}$ and its feature-based representations $\bm{y}$ exists, as specified in Proposition.\ref{Pro1}.

\begin{proposition}\label{Pro1}
For any two OHLC data $\bm{x}$ and $\tilde{\bm{x}}$ with their corresponding feature-based representations $\bm{y}$ and $\tilde{\bm{y}}$, then it holds $\bm{x}=\tilde{\bm{x}}$ only if $\bm{y}=\tilde{\bm{y}}$ holds.
\end{proposition}

Note that through feature information extraction, the feature-based representation $\bm{y}$ maps the original OHLC data $\bm{x}$ from the constricted $\mathbb{R}^4$ space with constraints in Definition.\ref{OHLC data} into the full $\mathbb{R}^4$ space, i.e.,
\vspace*{-0.5cm}
$$\mathbb{R}^{4}=\left\{\bm{y}=\left(y^{(1)}, y^{(2)}, y^{(3)}, y^{(4)}\right)^{\prime},-\infty<y^{(j)}<+\infty, j=1,2,3,4\right\}.$$
\vspace*{-1.5cm}

Here, we can directly perform vector linear and inner product operations based on classic linear algebra principles.
Furthermore, from Proposition.\ref{Pro1} and Definition.\ref{feature information}, we can easily formulate the inverse transformation from the feature-based representations $\bm{y}$ to the original OHLC data $\bm{x}$ according to Equation (\ref{x}):
\vspace*{-0.3cm}
\begin{equation}\label{x}
  \left\{\begin{array}{l}
  x^{(l)}=\exp \left\{y^{(1)}\right\} \\
  x^{(h)}=\exp \left\{y^{(1)}\right\}+\exp \left\{y^{(2)}\right\} \\
  x^{(o)}=\frac{1}{1+\exp \left\{y^{(3)}\right\}} x^{(l)}+\frac{\exp \left\{y^{(3)}\right\}}{1+\exp \left\{y^{(3)}\right\}} x^{(h)} \\
  x^{(c)}=\frac{1}{1+\exp \left\{y^{(4)}\right\}} x^{(l)}+\frac{\exp \left\{y^{(4)}\right\}}{1+\exp \left\{y^{(4)}\right\}} x^{(h)} \\
\end{array}\right..
\end{equation}

In summary, the novel feature-based representations not only eliminate the inherent constraints, providing convenience for further analysis but also are a one-to-one match with explicit inverse transformation in Equation (\ref{x}), which means that any analysis of the feature-based representations can be reversed to the original data. From this perspective, the proposed method may cast a new insight to investigate the OHLC data.

Finally, we show additional remarks on the assumption in Definition.\ref{feature information}. Note that assumptions $x^{(o)}, x^{(h)}, x^{(l)}$, and $x^{(c)}$ are not equal to each other $(\text{except for} \; x^{(o)} \equiv x^{(c)})$ implies that $x^{(h)} \neq x^{(l)} \neq 0$ and $\lambda^{(o)}, \lambda^{(c)} \notin \left\{ 0,1 \right\}$. These assumptions always hold under normal circumstances. However, when these assumptions are occasionally invalid, we recommend the following preprocessing procedure.
(1) When the subject is on trade suspension and all prices equal to $0$, namely, $x^{(o)}=x^{(h)}=x^{(l)}=x^{(c)}=0$, we exclude these extreme cases in the raw data.
(2) When $\lambda^{(o)}$ or $\lambda^{(c)}$ is equal to $0$, it corresponds to $x^{(o)}=x^{(l)}$ or $x^{(c)}=x^{(l)}$, respectively.
We add a random term to $x^{(o)}$ or $x^{(c)}$ and make $\lambda^{(o)}$ or $\lambda^{(c)}$ slightly greater than 0.
(3) When $\lambda^{(o)}$ or $\lambda^{(c)}$ is equal to $1$, it indicates that $x^{(o)}=x^{(h)}$ or $x^{(c)}=x^{(h)}$, respectively. We subtract a random term from $x^{(o)}$ or $x^{(c)}$ to make $\lambda^{(o)}$ or $\lambda^{(c)}$ slightly less than $1$.
(4) When certain subject reaches limit-up or limit-down as soon as the opening quotation, that is, $x^{(o)}=x^{(h)}=x^{(l)}=x^{(c)}\neq0$.
If limit-up (or limit-down) happens, we firstly multiply $x^{(c)}$ (or $x^{(o)})$ and $x^{(h)}$ by 1.1 to make a relatively large interval. And then conduct measurements given in circumstances (2) and (3).

\subsection{\textit{The sample version of Feature-based representations}} \label{Sec_2.2}

Let $\{\textbf{\textrm{x}}_i\}_{i=1}^n$ be independent and identically distributed (iid) samples with size $n$ from the $p$-dimensional vector of OHLC data $\textbf{\textrm{x}}=(\bm{x}_1,\cdots,\bm{x}_p)^{\prime}$, where $\bm{x}_j$ is the $j$th feature of OHLC data in Definition.\ref{OHLC data}. Denote $\textbf{\textrm{X}}$ as the $n\times p$ matrix of OHLC data with elements of the $i$th row being $\textbf{\textrm{x}}^\prime_i$ (i.e., $\textbf{\textrm{X}}=(\textbf{\textrm{x}}_1,\cdots,\textbf{\textrm{x}}_n)^\prime$). Rewrite $\textbf{\textrm{X}}$ in the form of column (i.e., $\textbf{\textrm{X}}=(\bm{X}_1,\cdots,\bm{X}_p)$), where the $j$th column $\bm{X}_j=(\bm{x}_{1j},\cdots,\bm{x}_{nj})^\prime$ is the $j$th sample feature of OHLC data.

Applying the proposed procedure in Definition.\ref{feature information} to each OHLC data $\bm{x}_{ij}$ yields the corresponding feature-based representation $\bm{y}_{ij}$. Correspondingly, let $\textbf{\textrm{Y}}$ be the $n\times p$ matrix of feature-based representation for OHLC data with elements of the $i$th row being $\textbf{\textrm{y}}^\prime_i$ (i.e., $\textbf{\textrm{Y}}=(\textbf{\textrm{y}}_1,\cdots,\textbf{\textrm{y}}_n)^\prime$). Rewrite $\textbf{\textrm{Y}}$ in the form of column (i.e., $\textbf{\textrm{Y}}=(\bm{Y}_1,\cdots,\bm{Y}_p)$), where the $j$th column $\bm{Y}_j=(\bm{y}_{1j},\cdots,\bm{y}_{nj})^\prime$ is the $j$th sample feature-based representations.

Next, we describe the vector space structure formed by the sample feature-based representations with size $n$, denoted by $\mathbb{R}^4_n$, together with its basic operations.
\begin{definition}\label{Rn4_space} (\textbf{Vector space structure of the sample feature-based representations with size $n$ and its basic operations}). Suppose that $\textbf{Y}=(\bm{y}_1,\cdots,\bm{y}_n)^\prime$ are derived from the iid sample feature of OHLC data matrix $\textbf{X}=(\bm{x}_1,\cdots,\bm{x}_n)^\prime$, we call the vector space formed by $\textbf{Y}$ as the space of sample feature-based representations with size $n$, that is:
\vspace*{-0.6cm}
\begin{equation*}
  \mathbb{R}_{n}^{4}=\left\{\bm{Y}=\left(\bm{y}_{1}, \bm{y}_{2}, \ldots, \bm{y}_{n}\right)^{\prime} ; \bm{y}_{i} \in \mathbb{R}^{4} \; (i=1,2, \ldots, n)\right\}.
\end{equation*}
\vspace*{-1.5cm}

For $\bm{Y}_j$, $\bm{Y}_k \in \mathbb{R}_n^{4}$ and $\beta\in\mathbb{R}$, we define the corresponding operators:
\vspace*{-0.2cm}
\begin{itemize}
  \item[](1) \textbf{Addition operator $\oplus$}:
  \vspace*{-0.5cm}
  \begin{equation}\label{Eq_plus}
  \boldsymbol{Y}_{j} \oplus \boldsymbol{Y}_{k}=\left(\boldsymbol{y}_{1 j}+\boldsymbol{y}_{1 k}, \boldsymbol{y}_{2 j}+\boldsymbol{y}_{2 k}, \ldots, \boldsymbol{y}_{n j}+\boldsymbol{y}_{n k}\right)^{\prime} \in \mathbb{R}_{n}^{4},
  \vspace*{-0.5cm}
  \end{equation}
  where $\forall i=1,2,\ldots,n$, $\bm{y}_{ij}+\bm{y}_{ik}=\left(y_{ij}^{(1)}+y_{ik}^{(1)}, y_{ij}^{(2)}+y_{ik}^{(2)}, y_{ij}^{(3)}+y_{ik}^{(3)}, y_{ij}^{(4)}+y_{ik}^{(4)}\right)^{\prime} \in \mathbb{R}^{4}$.
  \item[](2) \textbf{Scalar multiplication operator $\otimes$}:
  \vspace*{-0.5cm}
  \begin{equation}\label{Eq_times}
  \beta \otimes \bm{Y}_{j}=\left(\beta \bm{y}_{1j}, \beta \bm{y}_{2j}, \ldots, \beta \bm{y}_{nj}\right)^{\prime} \in \mathbb{R}_{n}^{4},
  \vspace*{-0.5cm}
  \end{equation}
  where $\forall i=1,2,\ldots,n$, $\beta \bm{y}_{ij}=\left(\beta y_{ij}^{(1)}, \beta y_{ij}^{(2)}, \beta y_{ij}^{(3)}, \beta y_{ij}^{(4)}\right)^{\prime} \in \mathbb{R}^{4}$.
  \item[](3) \textbf{Inner product operator $(\cdot, \cdot)_{\mathbb{R}_n^4}$}:
  \vspace*{-0.5cm}
  \begin{equation}\label{Eq_innerproduct}
  \left(\boldsymbol{Y}_{j}, \boldsymbol{Y}_{k}\right)_{\mathbb{R}_{n}^{4}}=\sum_{i=1}^{n}\left\langle\boldsymbol{y}_{i j}, \boldsymbol{y}_{i k}\right\rangle_{\mathbb{R}^{4}},
  \vspace*{-0.5cm}
  \end{equation}
  where $<\cdot, \cdot>_{\mathbb{R}^{4}}$ represents for the inner product operator in $\mathbb{R}^{4}$, which is defined as
  $\left\langle \bm{y}_{ij}, \bm{y}_{ik}\right\rangle_{\mathbb{R}^{4}}=\sum_{d=1}^{4} y_{ij}^{(d)} y_{ik}^{(d)}$.
\end{itemize}
\end{definition}
\vspace*{-0.2cm}

Furthermore, we can deduce the subtraction operator in $\mathbb{R}_{n}^{4}$ according to the definitions of addition and scalar multiplication operators, that is:
\vspace*{-0.5cm}
\begin{equation}\label{Eq_minus}
\bm{Y}_{j} \ominus \bm{Y}_{k}=\bm{Y}_{j} \oplus(-1) \otimes \bm{Y}_{k}=\left(\bm{y}_{1j}-\bm{y}_{1k}, \bm{y}_{2j}-\bm{y}_{2k}, \ldots, \bm{y}_{nj}-\bm{y}_{nk}\right)^{\prime},
\vspace*{-0.5cm}
\end{equation}
for $\forall i=1,2,\ldots,n$, $\bm{y}_{ij}-\bm{y}_{ik}=\left(y_{i j}^{(1)}-y_{i k}^{(1)}, y_{i j}^{(2)}-y_{i k}^{(2)}, y_{i j}^{(3)}-y_{i k}^{(3)}, y_{i j}^{(4)}-y_{i k}^{(4)}\right)^{\prime}$. The element zero $\bm{0}^{(0)}$ in $\mathbb{R}_{n}^{4}$ is:
$\mathbf{0}^{(0)}=(\mathbf{0}, \mathbf{0}, \ldots, \mathbf{0})^{\prime} \in \mathbb{R}_{n}^{4}$,
where $\mathbf{0}=(0,0,0,0)^{\prime} \in \mathbb{R}^{4}$.

The following standard properties hold, making them analogous to translation and scalar multiplication in real space.

\begin{proposition}\label{Proposition_2}
For any $\bm{Y}_j, \bm{Y}_k, \bm{Y}_l \in \mathbb{R}_n^{4}$ and $\beta \in \mathbb{R}$, the inner product operation in $\mathbb{R}_{n}^{4}$ satisfies:
\begin{itemize}
  \item[](1) \textbf{Non-negative property}: $\left(\boldsymbol{Y}_{j}, \boldsymbol{Y}_{j}\right)_{\mathbb{R}_{n}^{4}} \geq 0$, and $``="$ holds if and only if $\bm{Y}_j=\bm{0}^{(0)}$;
  \item[](2) \textbf{Commutative property}: $\left(\boldsymbol{Y}_{j}, \boldsymbol{Y}_{k}\right)_{\mathbb{R}_{n}^{4}}=\left(\boldsymbol{Y}_{k}, \boldsymbol{Y}_{j}\right)_{\mathbb{R}_{n}^{4}};$
  \item[](3) \textbf{Associative property}: $\left(\boldsymbol{Y}_{j}, \boldsymbol{Y}_{k} \oplus \boldsymbol{Y}_{l}\right)_{\mathbb{R}_{n}^{4}}=\left(\boldsymbol{Y}_{j}, \boldsymbol{Y}_{k}\right)_{\mathbb{R}_{n}^{4}}+\left(\boldsymbol{Y}_{j}, \boldsymbol{Y}_{l}\right)_{\mathbb{R}_{n}^{4}};$
  \item[](4) \textbf{Linear property}: $\left(\beta \otimes \bm{Y}_{j}, \bm{Y}_{k}\right)_{\mathbb{R}_{n}^{4}}=\beta\left(\bm{Y}_{j}, \boldsymbol{Y}_{k}\right)_{\mathbb{R}_{n}^{4}}.$
\end{itemize}
\end{proposition}

Conclusions (3) and (4) in Proposition.\ref{Proposition_2} will play an important role in the derivation of the subsequent pseudo-PCA approach. Before we present the main procedure, we will give some basic summary statistics based on the sample feature-based representations $\textbf{\textrm{Y}}=\{\bm{y}_{ij}\}_{i=1,\cdots,n; \; j=1,\cdots,p}$.

\begin{definition}\label{digital characteristics} (\textbf{Sample mean, variance and correlation coefficient for feature-based representations}). For any $\bm{Y}_j$, $\bm{Y}_k \in \mathbb{R}_n^{4}$ and $\beta\in\mathbb{R}$:
\begin{itemize}
  \item [](1) \textbf{Sample mean} for the $j$th feature-based representation is:
  \vspace*{-0.7cm}
  \begin{equation}\label{Eq_samplemean}
  \overline{\boldsymbol{Y}}_{j}=\frac{1}{n}\left(\boldsymbol{y}_{1 j}+\boldsymbol{y}_{2 j}+\cdots+\boldsymbol{y}_{n j}\right) \in \mathbb{R}^{4}.
  \vspace*{-0.7cm}
  \end{equation}
  \item[](2) \textbf{Sample covariance} for the $j$th and $k$th feature-based representations is:
  \vspace*{-0.5cm}
  \begin{equation}\label{Eq_samplecov}
  S_{j k}=\frac{1}{4n} \sum_{i=1}^{n}\left\langle\boldsymbol{y}_{i j}-\overline{\boldsymbol{Y}}_{j}, \; \boldsymbol{y}_{i k}-\overline{\boldsymbol{Y}}_{k}\right\rangle_{\mathbb{R}^{4}} \in \mathbb{R}.
  \vspace*{-0.5cm}
  \end{equation}
  $j=k$ yields the sample variance for the $j$th feature-based representation:
  \vspace*{-0.5cm}
  \begin{equation}\label{Eq_samplevar}
  S_{j}^{2}=\frac{1}{4n} \sum_{i=1}^{n}\left\langle\boldsymbol{y}_{i j}-\overline{\boldsymbol{Y}}_{j}, \; \boldsymbol{y}_{i j}-\overline{\boldsymbol{Y}}_{j}\right\rangle_{\mathbb{R}^{4}} \in \mathbb{R}.
  \vspace*{-0.5cm}
  \end{equation}
  \item[](4) \textbf{Sample correlation coefficient} for the $j$th and $k$th feature-based representations is:
  \vspace*{-0.5cm}
  \begin{equation}\label{Eq_samplecorr}
  r_{j k}=\frac{S_{j k}}{S_{j} S_{k}} \in \mathbb{R}.
  \vspace*{-0.5cm}
  \end{equation}
\end{itemize}
\end{definition}
The sample variance-covariance matrix $\bm{\Sigma}$ and correlation coefficient matrix $\bm{W}$ for the feature-based representation matrix $\textbf{Y}$ are:
\begin{equation}\label{Eq 22}
\bm{\Sigma}=\left(\begin{array}{ccc}
S_{1}^{2} & \cdots & S_{1 p} \\
\vdots & \ddots & \vdots \\
S_{p 1} & \cdots & S_{p}^{2}
\end{array}\right) \in \mathbb{R}^{p \times p},\ \ \bm{W}=\left(\begin{array}{ccc}
1 & \cdots & r_{1 p} \\
\vdots & \ddots & \vdots \\
r_{p 1} & \cdots & 1
\end{array}\right) \in \mathbb{R}^{p \times p}.
\end{equation}

It is important to note that although each element in the feature information matrix $\textbf{\textrm{Y}}=\{\bm{y}_{ij}\}_{i=1,\cdots,n; \; j=1,\cdots,p}$ belongs to $\mathbb{R}^4$ space, its variance-covariance matrix $\bm{\Sigma}$ and correlation coefficient matrix $\bm{W}$ are $p \times p$ dimensional matrices in the real space. This conclusion will make it convenient to derive the pseudo-PCA for OHLC data in $\mathbb{R}_n^4$ space. In addition, we can also standardize the elements in matrix $\textbf{\textrm{Y}}$ according to
\vspace*{-0.5cm}
\begin{equation}\label{Eq 24}
\boldsymbol{y}_{i j}^{*}=\frac{1}{S_{j}}\left(\boldsymbol{y}_{i j}-\overline{\boldsymbol{Y}}_{j}\right) \in \mathbb{R}^{4}.
\vspace*{-0.5cm}
\end{equation}

\vspace*{-0.8cm}
\section{Pseudo-PCA procedure} \label{Sec 3}

In multivariate statistics, PCA is a widely used method of dimension reduction.
Like the objective of the classic PCA, for the feature information variables $\left \{ \bm{Y}_j \in \mathbb{R}_{n}^{4}, \; j=1,2,\ldots,p \right \}$ of OHLC data in $\mathbb{R}_n^4$ space, the aim of pseudo-PCA is also to reduce the $p$-dimensional space to $m$-dimensional $\left( m \leq p \right )$ under the premise of minimal the loss of the variance information.
To clarify, it uses the maximum variation direction of the feature information to find a set of so-called ¡°pseudo-PCs¡±, denoted as $\boldsymbol{F}_{h}=\left(\boldsymbol{f}_{1 h}, \boldsymbol{f}_{2 h}, \ldots, \boldsymbol{f}_{n h}\right)^{\prime} \in \mathbb{R}_{n}^{4} \; (h=1, \ldots, m, m \leq p)$, where each pseudo-PC $\bm{F}_h$ is a linear combination of $\bm{Y}_1, \bm{Y}_2,\ldots, \bm{Y}_p$ with the loading coefficients being $\bm{u}_h=(u_{h1},\cdots,u_{hp})^\prime$, namely:
\vspace*{-0.6cm}
\begin{equation}\label{Eq 25}
\boldsymbol{F}_{h}=u_{h 1} \otimes \boldsymbol{Y}_{1} \oplus u_{h 2} \otimes \boldsymbol{Y}_{2} \oplus \cdots \oplus u_{h p} \otimes \boldsymbol{Y}_{p}.
\vspace*{-0.5cm}
\end{equation}

The pseudo-PCA of OHLC data expects $\sum_{h=1}^{m} \operatorname{Var}\left(\boldsymbol{F}_{h}\right)$ to get the maximum value, and $\operatorname{Var}\left(\boldsymbol{F}_{1}\right) \geq \operatorname{Var}\left(\boldsymbol{F}_{2}\right) \geq \cdots \geq \operatorname{Var}\left(\boldsymbol{F}_{m}\right)$.
In addition, the combination coefficients $\boldsymbol{u}_{h}=\left(u_{h 1}, u_{h 2}, \ldots, u_{h p}\right)^{\prime} \in \mathbb{R}^{p} \; (h=1,2, \ldots, m)$ should be orthonormal and normalized.

Assuming that $\bm{Y}_1, \bm{Y}_2,\ldots, \bm{Y}_p$ have been standardized, there is $\overline{\boldsymbol{Y}}_{j}=\mathbf{0} \in \mathbb{R}^{4} \; (j=1,2, \ldots, p)$, and
$\overline{\boldsymbol{F}}_{h}=\mathbf{0} \in \mathbb{R}^{4}$. Therefore, using the Proposition.\ref{Proposition_2} and the Definition.\ref{digital characteristics}, the sample variance of $\bm{F}_h$ is:
\vspace*{-0.5cm}
\begin{equation}\label{Eq 26}
\operatorname{Var}\left(\boldsymbol{F}_{h}\right)=\frac{1}{n}\left(\boldsymbol{F}_{h}, \boldsymbol{F}_{h}\right)_{\mathbb{R}_{n}^{4}}
=\frac{1}{n}\left(u_{h 1} \otimes \boldsymbol{Y}_{1} \oplus \cdots \oplus u_{h p} \otimes \boldsymbol{Y}_{p}, u_{h 1} \otimes \boldsymbol{Y}_{1} \oplus \cdots \oplus u_{h p} \otimes \boldsymbol{Y}_{p}\right)_{\mathbb{R}_{n}^{4}}=\boldsymbol{u}_{h}^{\prime} \boldsymbol{W u}_{h}.
\end{equation}

In Equation (\ref{Eq 26}), $\bm{W}$ is the correlation coefficient matrix of the feature information variable set $\{\bm{Y}_1, \bm{Y}_2,\ldots, \bm{Y}_p\}$, which is a $p \times p$ dimension matrix in the real number domain. Thus, we get Theorem.\ref{Theorem_1}.
\newtheorem{theorem}{\bf Theorem}
\begin{theorem}\label{Theorem_1}
If $\bm{W}$ is the correlation coefficient matrix of the feature information variable set $\{\bm{Y}_1, \bm{Y}_2,\ldots, \bm{Y}_p \; | \; \bm{Y}_j \in  \mathbb{R}_n^4, \; j=1,2,\ldots,p \} $, and $\boldsymbol{u}_{h}=\left(u_{h 1}, u_{h 2}, \ldots, u_{h p}\right)^{\prime} \in \mathbb{R}^{p} \; (h=1,2, \ldots, m, \; m \leq p)$, remark the pseudo-principle component as $\boldsymbol{F}_{h}=u_{h 1} \otimes \boldsymbol{Y}_{1} \oplus u_{h 2} \otimes \boldsymbol{Y}_{2} \oplus \cdots \oplus u_{h p} \otimes \boldsymbol{Y}_{p}$.
Under the condition of $\boldsymbol{u}_{h}$ is orthonormal and normalized, deriving the first $m$ pseudo-PCs is equivalent to solving the following optimization problem:

\vspace{5pt}
\begin{centering}
$\max \sum_{h=1}^{m} \boldsymbol{u}_{h}^{\prime} \boldsymbol{W u}_{h}$ \\
\vspace{-40pt}
$$ \emph{\text{s.t.}} \begin{cases}
    \boldsymbol{u}_{h}^{\prime} \boldsymbol{u}_{k}=
    \begin{cases}
    \begin{array}{ll}
        1 & h=k \\
        0 & h \neq k
        \end{array}, \quad h, k=1,2, \ldots, m, \; m \leq p
    \end{cases}\\
    \bm{u}_1' \bm{W} \bm{u}_1 \geq \bm{u}_2' \bm{W} \bm{u}_2 \geq \cdots \geq \bm{u}_m' \bm{W} \bm{u}_m, \; m \leq p
\end{cases}$$
\end{centering}
\end{theorem}
\vspace*{-0.3cm}

According to linear algebra theory, conducting eigen-decomposition of matrix $\bm{W}$ and deriving the first $m$ eigenvalues $\lambda_1 \geq \lambda_2 \geq \cdots \geq \lambda_m > 0$, the corresponding eigenvectors $\bm{u}_1, \bm{u}_2,..., \bm{u}_m$ are the solutions of the optimization problem in Theorem \ref{Theorem_1} \citep{abdi2010principal}.
Then, the first $m$ pseudo-principle components $\bm{F}_{1},\bm{F}_{2},...,\bm{F}_{m}$ can be directly obtained by Equation (\ref{Eq 25}).

In addition, it is not difficult to prove that after the pseudo-PCA, the following important properties are established. The corresponding proof can refer to Appendix B.
\begin{proposition}\label{Proposition_3}
If the feature information variables $\bm{Y}_1, \bm{Y}_2,\ldots, \bm{Y}_p$ are standardized, the first $m \; (m \leq p)$ pseudo-principle components $\bm{F}_h \in \mathbb{R}_n^4 \; ( 1 \leq h \leq m)$ derived from the pseudo-PCA satisfy the following properties:\\
(1)\; The sample mean of $\bm{F}_h$ is equal to $0$, that is, $\overline{\boldsymbol{F}}_{h}=\mathbf{0} \in \mathbb{R}^{4}$; \\
(2)\; The sample variance of $\bm{F}_h$ is equal to $\lambda_h$, namely, $\operatorname{Var}\left(\boldsymbol{F}_{h}\right)=\lambda_{h}$; \\
(3)\; For any $\bm{F}_j, \bm{F}_k \in \mathbb{R}_n^4 ( 1 \leq j,k \leq p)$, if $j \neq k$, their sample covariance $Cov(\bm{F}_j, \bm{F}_k) = 0$; \\
(4)\; $\sum_{h=1}^{p} Var(\bm{F}_h)=p.$
\end{proposition}

Like the classical PCA, we can define the cumulative contribution rate $Q_m$ based on the conclusions (2) and (4) in Proposition.2 as follows:
\vspace*{-0.5cm}
\begin{equation}\label{Eq 27}
Q_m=\frac{1}{p}\sum_{h=1}^{m}\lambda_h.
\vspace*{-0.5cm}
\end{equation}
There is $0 < Q_m \leq 1$.
A larger $Q_m$ indicates that the pseudo-principle components retain more variance information after conducting pseudo-PCA, and the analysis accuracy is higher.

\begin{figure}[!h]
  \centering
  \includegraphics[scale=0.38]{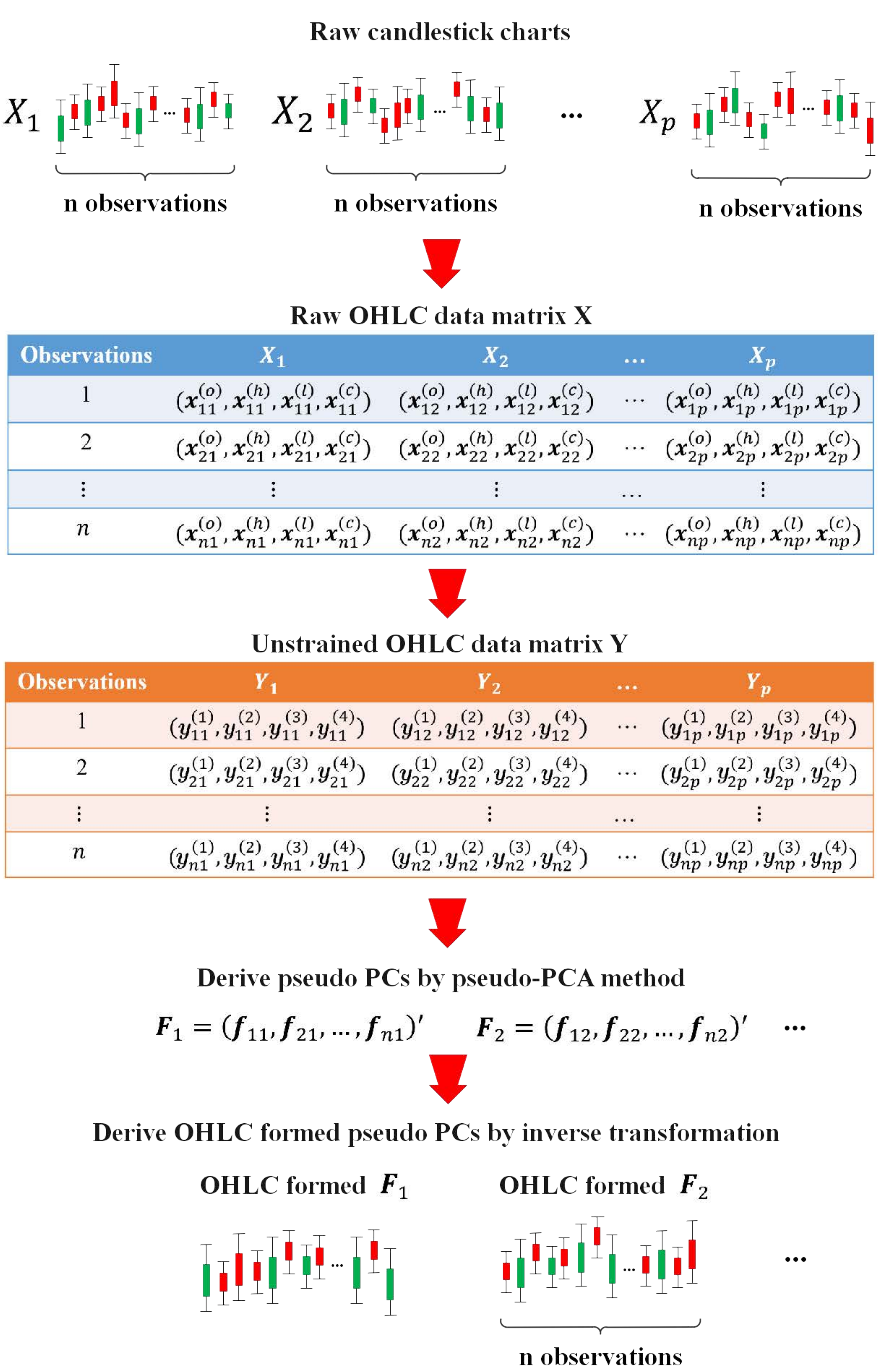}
  \caption{The pseudo-PCA modelling framework for OHLC data}\label{Fig process}
\end{figure}

To summarize, a pseudo-PCA modeling framework is illustrated by Figure.\ref{Fig process} and the specific implementation process is summarized as Algorithm.\ref{pca-ohlc}.

\begin{algorithm}[!htb]
\setstretch{2}
	\caption{$\;$ Pseudo-PCA procedure for OHLC data}
	\label{pca-ohlc}
	\begin{algorithmic}[1]
		\Require $\textrm{\textbf{X}}=(\bm{X}_1,\cdots,\bm{X}_p)$  \Comment{$n$ samples from the $p$-dimensional vector of OHLC data}
		\Ensure  $\textrm{\textbf{F}}=(\bm{F}_1,\cdots,\bm{F}_p)$ \Comment{$\bm{F}_j$ is the $j$th pseudo-PC}
        \State Calculate the matrix of feature-based representations $\textrm{\textbf{Y}}$ according to Equation (\ref{Eq_1}) and (\ref{Eq_2}). 	
		\State Derive standardized $\textrm{\textbf{Y}}$ according to Equation (\ref{Eq 24}). For the sake of convenience, still record the standardized matrix as $\textrm{\textbf{Y}}$.
		\State Calculate the correlation coefficient matrix $\bm{W} \in \mathbb{R}^{p \times p}$ of the standardized matrix $\textrm{\textbf{Y}}$.
        For the matrix $\bm{W}$, derive its first $m$ eigenvalues $\lambda_1 \geq \lambda_2 \geq \cdots \geq \lambda_m > 0$, and the corresponding eigenvectors $\bm{u}_1, \bm{u}_2,..., \bm{u}_m$.
        According to Equation (\ref{Eq 25}), the first $m$ pseudo-PCs $\bm{F}_{1},\bm{F}_{2},...,\bm{F}_{m}$ can be calculated.
		\State After obtaining $\boldsymbol{F}_{h}=\left(\boldsymbol{f}_{1 h}, \boldsymbol{f}_{2 h}, \ldots, \boldsymbol{f}_{n h}\right)^{\prime} \in \mathbb{R}_{n}^{4} \; (h=1,2, \ldots, m)$, since each element$\boldsymbol{f}_{i h}=\left(f_{i h}^{(1)}, f_{i h}^{(2)}, f_{i h}^{(3)}, f_{i h}^{(4)}\right)^{\prime} \in \mathbb{R}^{4} \; (1 \leq i \leq n)$, the corresponding OHLC formed pseudo-principle components $\textrm{\textbf{F}}$ can be obtained through Equation (\ref{x}).	
	\end{algorithmic}
\end{algorithm}
\vspace*{-0.5cm}

Finally, we realize the pseudo-PCA for OHLC data by extracting its feature information and successfully reduce the dimensionality of an OHLC data set with $n$ observations and $p$ variables to a new OHLC data set with $n$ observations and $m$ new components (pseudo-PCs), and $m \leq p$.
In this process, the variance loss of the feature information matrix is minimal.

\section{Simulations} \label{Sec simulation}
\vspace*{-0.3cm}

The performance of pseudo-PCA is evaluated via finite sample simulation studies in term of the following two aspects: (1) the ability to remove redundant variables and reduce the dimensionality of raw data, and (2) the consistency of the estimated loadings of pseudo-PCs.

Here we assume that $p=6$ and the unconstrained OHLC data variable vector $\textrm{\textbf{Y}}=( \bm{Y}_1, \bm{Y}_2, \bm{Y}_3, \bm{Y}_4 ,\bm{Y}_5,\bm{Y}_6)$ satisfy that (1) each the unconstrained OHLC data variable $\bm{Y}_j$ is mean zero for $j=1,2,\cdots,6$; and (2) the correlation coefficient matrix $\boldsymbol{W_{\mathrm{Y}}}$ is set as following: \begin{equation}\label{W}
\boldsymbol{W_{\mathrm{Y}}}=\left(\begin{array}{cccccc}
1 & \frac{0.5^{2}}{2} & \frac{0.5^{2}}{3} & \frac{0.5^{2}}{4} & \frac{1.25}{\sqrt{5.5}} & \frac{0.5}{\sqrt{25.5}} \\
\frac{0.5^{2}}{2} & 1 & \frac{0.5^{2}}{6} & \frac{0.5^{2}}{8} & \frac{4.25}{2 \sqrt{5.5}} & \frac{0.5}{2 \sqrt{25.5}} \\
\frac{0.5^{2}}{3} & \frac{0.5^{2}}{6} & 1 & \frac{0.5^{2}}{12} & \frac{0.5}{3 \sqrt{5.5}} & \frac{9.25}{3 \sqrt{25.5}} \\
\frac{0.5^{2}}{4} & \frac{0.5^{2}}{8} & \frac{0.5^{2}}{12} & 1 & \frac{0.5}{4 \sqrt{5.5}} & \frac{16.25}{4 \sqrt{25.5}} \\
\frac{1.25}{\sqrt{5.5}} & \frac{4.25}{2 \sqrt{5.5}} & \frac{0.5}{3 \sqrt{5.5}} & \frac{0.5}{4 \sqrt{5.5}} & 1 & \frac{2}{\sqrt{140.25}} \\
\frac{0.5}{\sqrt{25.5}} & \frac{0.5}{2 \sqrt{25.5}} & \frac{9.25}{3 \sqrt{25.5}} & \frac{16.25}{4 \sqrt{25.5}} & \frac{2}{\sqrt{140.25}} & 1
\end{array}\right).
\vspace*{-0.1cm}
\end{equation}
Actually, from Equation (\ref{W}), there are two redundant variables $\bm{Y}_5$ and $\bm{Y}_6$, i.e. $\bm{Y}_5 = \bm{Y}_1+\bm{Y}_2$ and $\bm{Y}_6 = \bm{Y}_3+\bm{Y}_4$. The sample sizes are $n=50,100,150,200$ with each case $300$ repeats, and we summary the performance of proposed procedure in term of two measurements: (1) the cumulative variance contribution rate of the first four pseudo-PCs;  and (2) the mean absolute percentage error (MAPE) of the eigenvectors, i.e.,  $\operatorname{MAPE}_j=\frac{\left\|\widehat{\boldsymbol{u}}_{j}-\boldsymbol{u}_{j}\right\|}{\left\|\boldsymbol{u}_{j}\right\|} \times 100 \%$ $(1 \leq j \leq 6)$. For each criteria, we also report the empirical standard derivations.

\begin{figure}[!htb]
  \centering
  \includegraphics[scale=0.6]{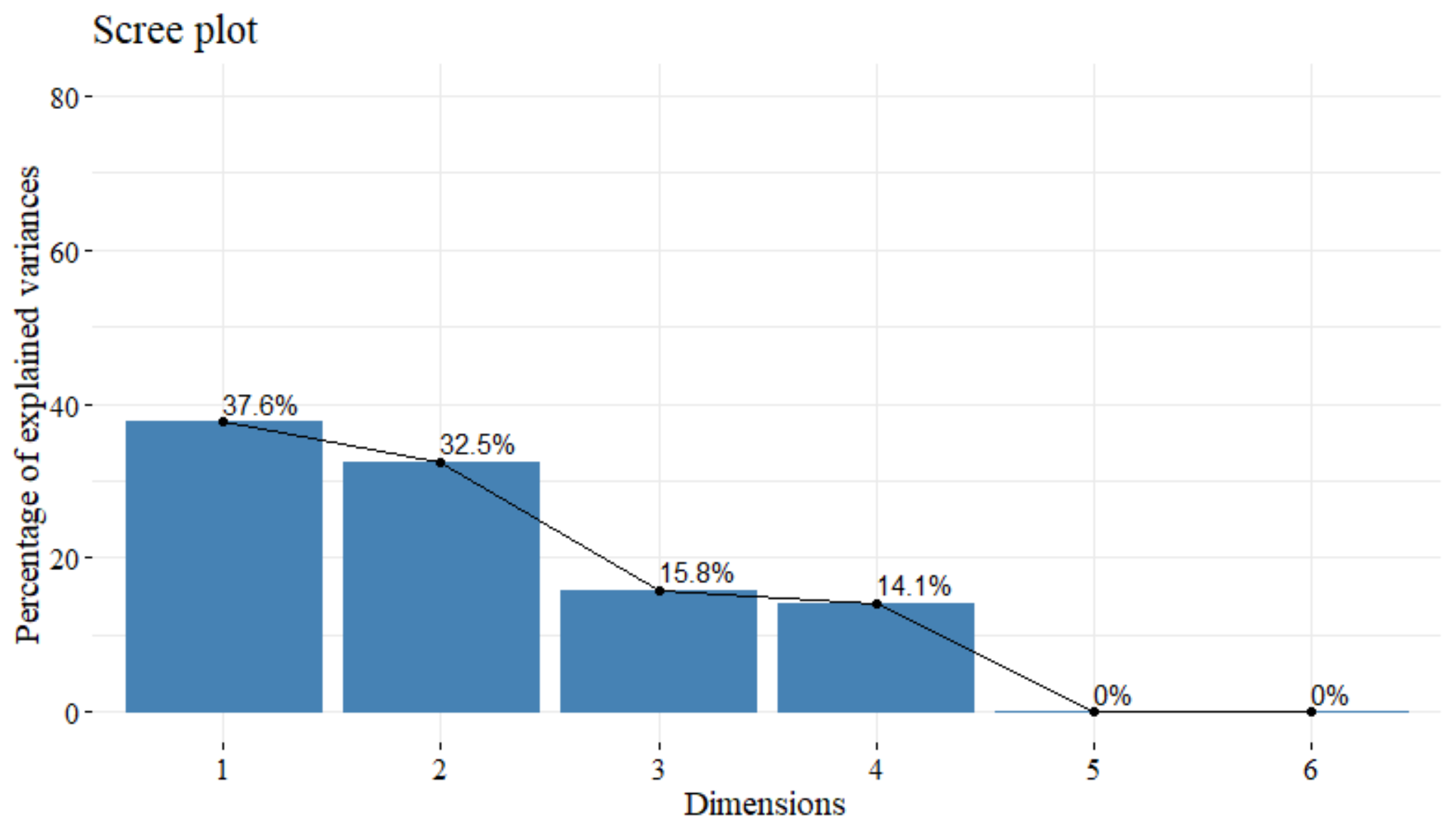}
  \caption{Cumulative variance contribution plot of simulation with $n=200$}
  \label{sim_cum}
\end{figure}

The cumulative variance contribution rate of the first four pseudo-PCs is very stable with the change of sample size and always reaches 100.0\%.
Take $n=200$ as an example, Figure.\ref{sim_cum} shows the 100.0\% explanatory capability of the first four pseudo-PCs.
This verified the extraordinary ability of pseudo-PCA to remove redundant variables.
Further, Table.\ref{Results} shows the results of MAPE and its empirical standard derivations with different sample size.
The estimation accuracy of $\textbf{PC5}$ and $\textbf{PC6}$ is relatively high and that of $\textbf{PC3}$ and $\textbf{PC4}$ is relatively low.
The mean values of MAPE range from 13.2\% to 27.9\%, and its empirical standard derivations varies from 0.136 to 0.269, which are both within the acceptable range.
Meanwhile, with the increase of sample size, MAPE and its empirical standard derivations show a decreasing trend, indicating that the estimated eigenvectors gradually converges to the theoretical values.
These results identified the consistency between the estimated loadings of pseudo-PCs derived from the sample data and the theoretical values.

\begin{table}[!h]
  \caption {Average of MAPE and its empirical standard deviation based on 300 repetitions}
  \label{Results}
  \renewcommand{\arraystretch}{1.5}
  \begin{center}
  \begin{tabular}{cccccccc}
  \hline \textbf{Sample size} & \textbf{PC1} & \textbf{PC2} & \textbf{PC3} & \textbf{PC4} & \textbf{PC5} & \textbf{PC6} & \textbf{Mean} \\
  \hline \multirow{2}{*} {$\mathrm{n}=50$} & 26.5\% & 28.5\% & 46.9\% & 47.1\% & 8.9\% & 9.3\% & 27.9\% \\
  & (0.308) & (0.301) & (0.320) & (0.318) & (0.185) & (0.184) & (0.269) \\
  \hline \multirow{2}{*} {$\mathrm{n}=100$} & 14.1\% & 15.2\% & 34.8\% & 34.6\% & 8.2\% & 8.4\% & 19.2\% \\
  & (0.162) & (0.159) & (0.259) & (0.260) & (0.197) & (0.196) & (0.206) \\
  \hline \multirow{2}{*} {$\mathrm{n}=150$} & 11.3\% & 12.6\% & 31.5\% & 31.6\% & 6.2\% & 6.4\% & 16.6\% \\
  & (0.141) & (0.139) & (0.226) & (0.225) & (0.163) & (0.162) & (0.176) \\
  \hline \multirow{2}{*} {$\mathrm{n}=200$} & 8.9\% & 9.8\% & 25.3\% & 25.2\% & 5.0\% & 5.1\% & 13.2\% \\
  & (0.067) & (0.066) & (0.204) & (0.205) & (0.138) & (0.137) & (0.136) \\
  \hline
  \end{tabular}
  \end{center}
\end{table}

\vspace*{-1.5cm}
\section{Empirical analysis} \label{Sec 4}
\vspace*{-0.2cm}

In order to illustrate the effectiveness of the proposed pseudo-PCA, this paper extracts the OHLC data of he spot prices of $6$ common foods of $20$ important agricultural product markets in China from Wind {\color{red}(\url{https://www.wind.com.cn/})}.
The $6$ OHLC data variables include: beef $(CNY/kg)$, lamb $(CNY/kg)$, pork $(CNY/kg)$, cucumber $(CNY/kg)$, potato $(CNY/kg)$, and onion $(CNY/kg)$.
The observed dates range from $1/1/2019$ to $31/12/2019$, and we use the first price recorded in the time period as the open price, last price recorded as the close price, and highest/lowest price that appears during the time period as the high/low price.

\begin{figure}[!htb]
  \centering
  \includegraphics[scale=0.7]{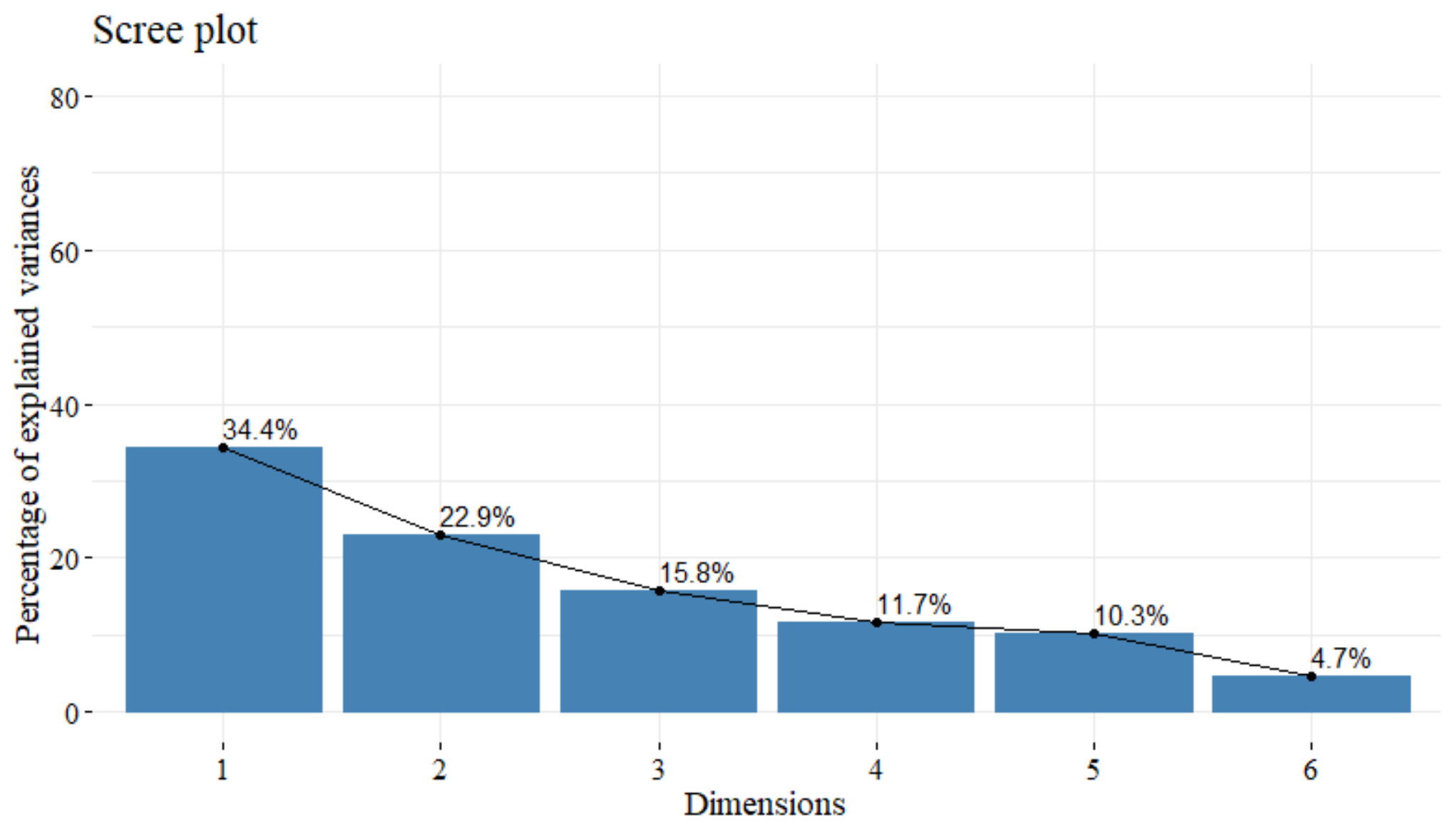}
  \caption{Cumulative variance contribution plot}
  \label{variance cumulative contribution plot}
\end{figure}

The original OHLC data matrix $\textrm{\textbf{X}}$ can be referred to Table.\ref{Raw OHLC data} in Appendix.B.
First, the standardized feature information matrix $\textrm{\textbf{Y}}$ is derived and shown in Table.\ref{Feature information data} in Appendix.B.
Conduct pseudo-PCA modelling on the standardized feature information matrix $\textrm{\textbf{Y}}$, the corresponding eigenvalues (EV), variance contribution rate (VCR), and cumulative contribution to the variance (CVCR) of the $6$ extracted pseudo-PCs are shown in Table.2.
The cumulative contribution rate to the variance of the first two pseudo-PCs reaches 57.4$\%$, and that of the first three pseudo-PCs reaches 73.2$\%$. This implies that the pseudo-principle components represent the information of the original data matrix well.
The corresponding variance cumulative contribution plot is shown in Figure.\ref{variance cumulative contribution plot}.
\vspace*{0.3cm}

\begin{table}[!htb]
  \label{VCR}
  \begin{spacing}{2}
  \begin{center}
  \setlength\tabcolsep{8 pt}
  \renewcommand{\arraystretch}{0.8}
  \caption {The EV, VCR and CVCR of the 6 pseudo-PCs}
  \begin{tabular}{cccc}
  \hline
  \textbf{pseudo-PCs} & \textbf{EV} & \textbf{VCR} & \textbf{CVCR} \\ \hline
  $\textbf{PC1}$                            & 2.065       & 0.344        & 0.344         \\ \hline
  $\textbf{PC2}$                             & 1.376       & 0.229        & 0.574         \\ \hline
  $\textbf{PC3}$                            & 0.950       & 0.158        & 0.732         \\ \hline
  $\textbf{PC4}$                             & 0.705       & 0.118        & 0.849         \\ \hline
  $\textbf{PC5}$                             & 0.619       & 0.103        & 0.953         \\ \hline
  $\textbf{PC6}$                             & 0.285       & 0.048        & 1.000         \\ \hline
  \end{tabular}
  \end{center}
  \end{spacing}
  \vspace*{-1.2cm}
\end{table}

Furthermore, the loading matrix of the first two pseudo-PCs is shown in Table.3.
The expressions of these two pseudo-PCs are:
\vspace*{-0.7cm}
\begin{eqnarray*}
\textbf{PC1}&=&0.222 \otimes \bm{Y}_{1} \oplus 0.115 \otimes \bm{Y}_{2} \oplus 0.248 \otimes \bm{Y}_{3} \oplus 0.543 \otimes \bm{Y}_{4} \oplus 0.599 \otimes \bm{Y}_{5} \oplus 0.471 \otimes \bm{Y}_{6}\\
\textbf{PC2}&=&0.563 \otimes \bm{Y}_{1} \oplus 0.687 \otimes \bm{Y}_{2} \oplus 0.330 \otimes \bm{Y}_{3} \ominus 0.199 \otimes \bm{Y}_{4} \ominus 0.129 \otimes \bm{Y}_{5} \ominus 0.213 \otimes \bm{Y}_{6}
\end{eqnarray*}

\begin{table}[!htb]
  \label{Loadings}
  \begin{spacing}{2}
  \begin{center}
  \vspace*{0.5cm}
  \setlength\tabcolsep{8 pt}
  \renewcommand{\arraystretch}{0.8}
  \caption {The loading matrix of the first two pseudo-PCs}
  \begin{tabular}{ccc}
  \hline
  \textbf{Feature information variables} & \textbf{PC1} & \textbf{PC2} \\ \hline
  Beef \; ($\bm{Y}_1$)                                   & 0.222        & 0.563        \\ \hline
  Lamb \; ($\bm{Y}_2$)                                    & 0.115        & 0.687        \\ \hline
  Pork \; ($\bm{Y}_3$)                                    & 0.248        & 0.330        \\ \hline
  Cucumber \; ($\bm{Y}_4$)                                & 0.543        & -0.199       \\ \hline
  Potato   \; ($\bm{Y}_5$)                                & 0.599        & -0.129       \\ \hline
  Onion    \; ($\bm{Y}_6$)                                 & 0.471        & -0.213       \\ \hline
  \end{tabular}
  \end{center}
  \end{spacing}
  \vspace*{-0.7cm}
\end{table}

\begin{figure}[!htb]
  \centering
  \includegraphics[scale=0.6]{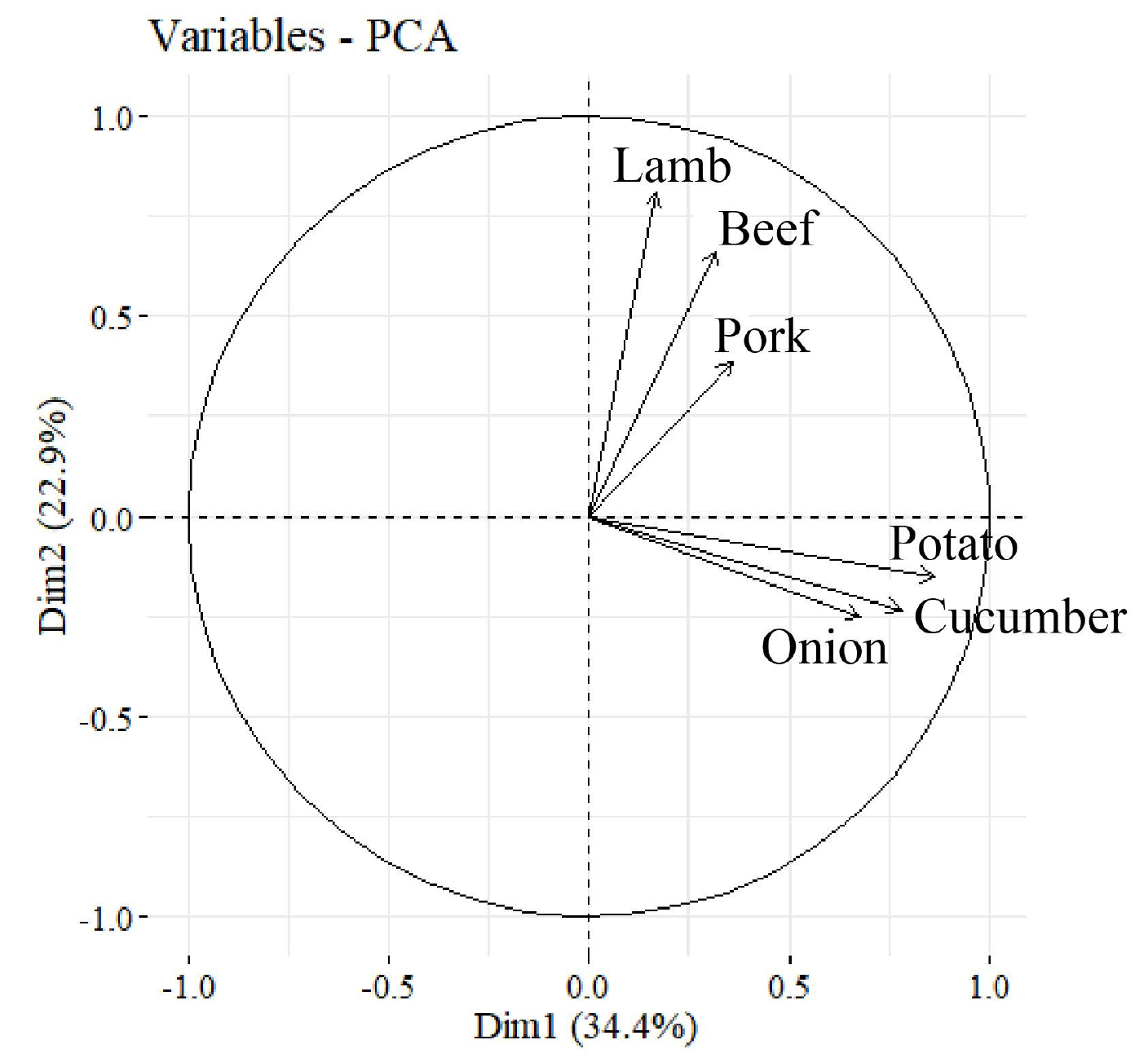}
  \caption{Correlation of the first two pseudo-PCs and the feature information variables}
  \label{Loading plot}
\end{figure}

\vspace*{-0.7cm}
The correlation between the first two pseudo-principle components and feature information variables are clearly visualized through Figure.\ref{Loading plot}.
In Figure.\ref{Loading plot}, positively related variables are close to each other, while negatively related variables are divergent. The lengths of the arrows represent the information quality of different variables.
It is observed that the loading factors' absolute values of the $\textbf{PC1}$ on cucumbers, potatoes, and onions are relatively large, which can be called a `Vegetable' factor.
The loading factors' absolute values of the $\textbf{PC2}$ on beef, lamb, and pork are comparatively large, and it can be called a `Meat' factor.

\begin{figure}[!htb]
  \centering
  \resizebox{16cm}{8.5cm}{\includegraphics{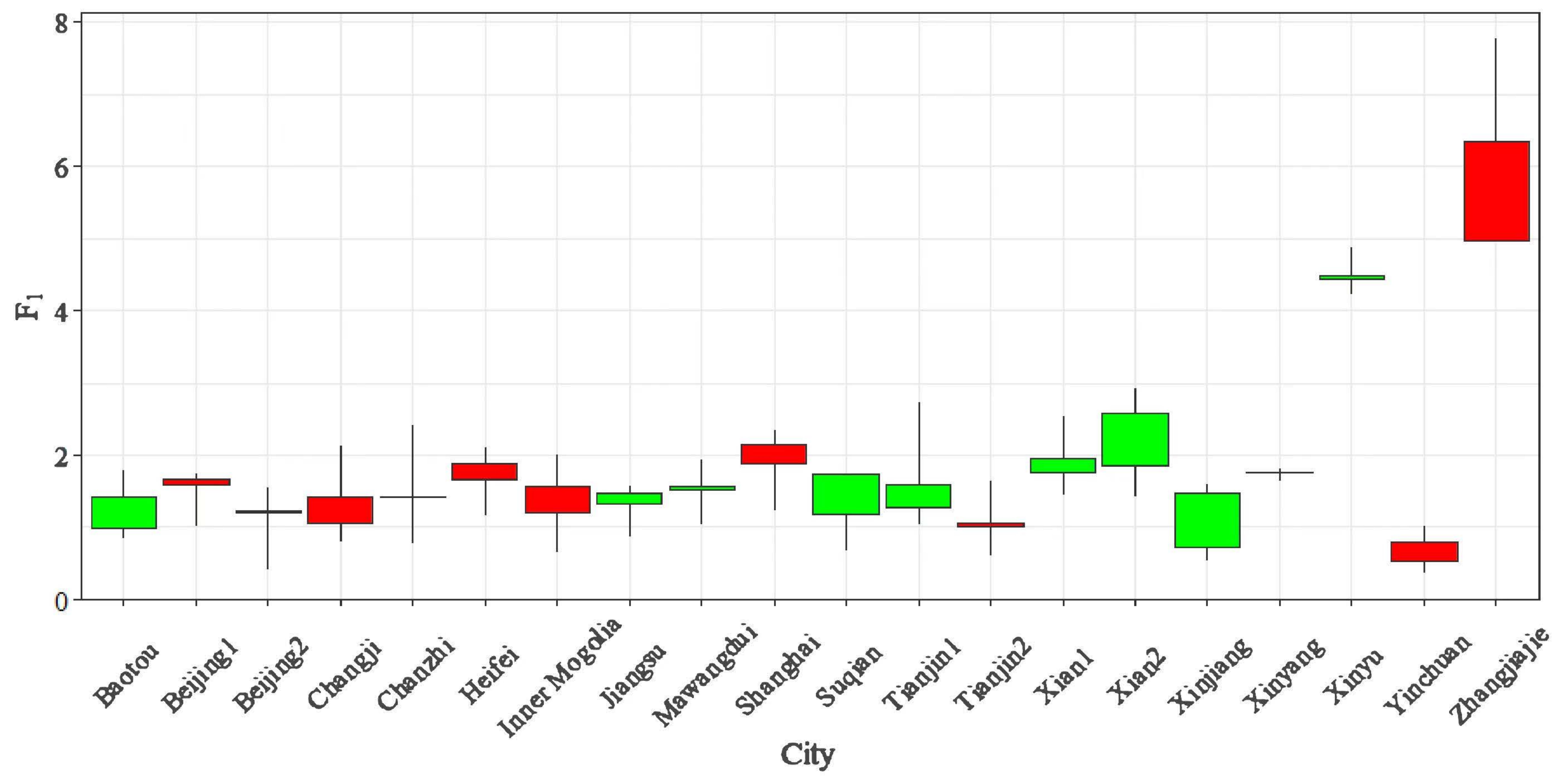}}
  \caption{The first OHLC formed pseudo-PC scores}\label{F1}
\end{figure}

\begin{figure}[!htb]
  \centering
  \resizebox{16cm}{8.5cm}{\includegraphics{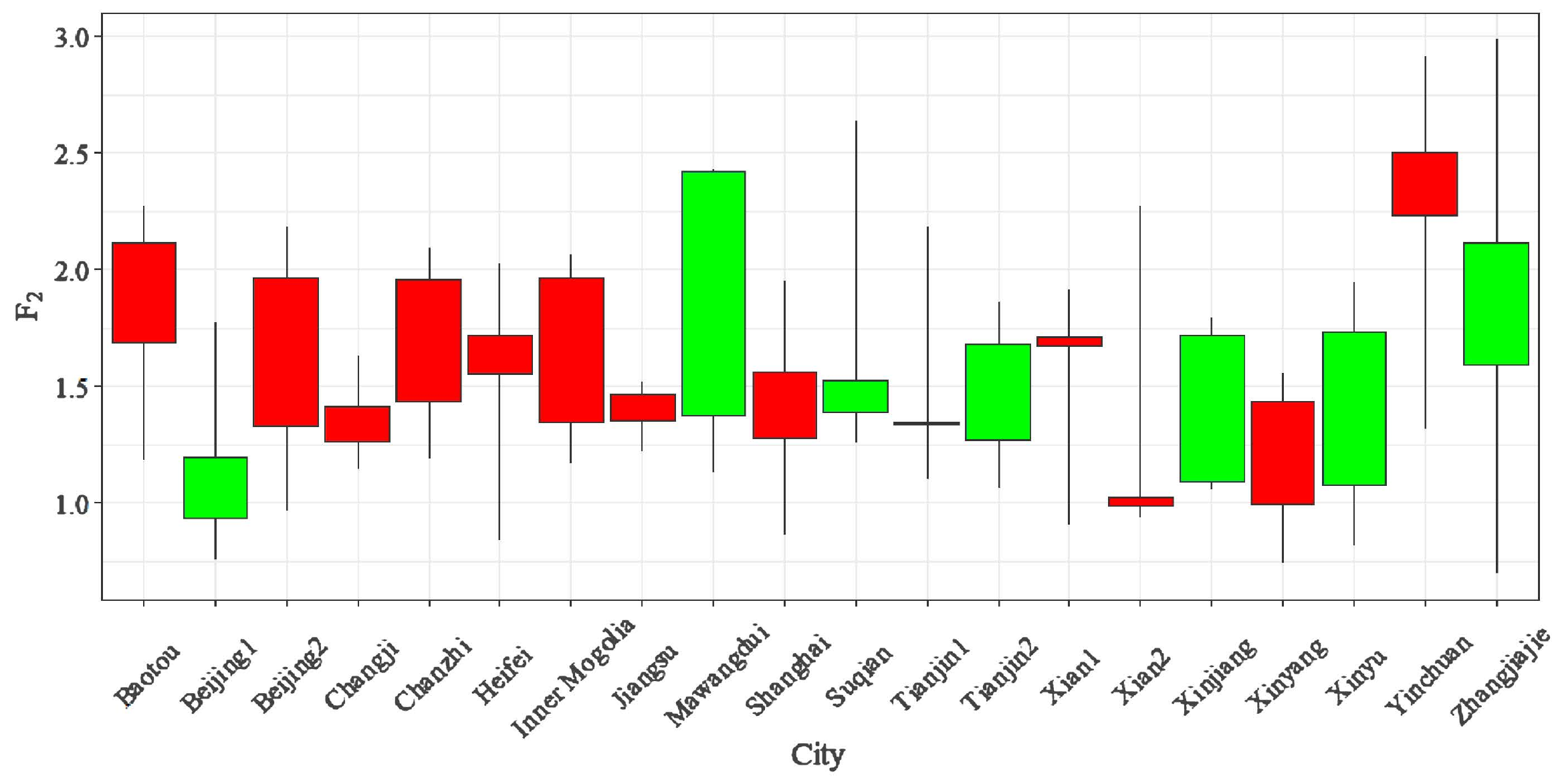}}
  \caption{The second OHLC formed pseudo-PC scores}\label{F2}
\end{figure}

According to Equation (\ref{Eq 25}), the scores of the first two pseudo-PCs can be calculated, and the corresponding OHLC formed data can be derived by using Equation (\ref{x}).
The results are shown in Figure.\ref{F1} and Figure.\ref{F2}, where the abbreviations of the 20 markets are used (the corresponding full name can be referred to Table.\ref{Market full name-abbreviated comparison table} in Abbreviation.C).

Based on the OHLC formed pseudo-PC scores, we can observe the size and fluctuation of the original data during the observation period.
For instance,
\textbf{(1) For the OHLC data converted from the first pseudo-PC scores (`Vegetable' factor)}, it is observed that the open, high, low, and open prices of Zhangjiajie and Xinyu have always been at a high level. This means that the prices of cucumbers, potatoes, and onions in these two cities maintained a high level through 2019.
At the same time, the difference between the low price and high price of Zhangjiajie is relatively large, indicating that the prices of cucumbers, potatoes, and onions in Zhangjiajie experienced great fluctuations in 2019.
While the difference between the upper and lower bounds of Xinyang is very small, corresponding to the prices of `Vegetable' in Xinyang were very stable in 2019.
Markets with green-body OHLC data had a relatively low comprehensive price for cucumbers, potatoes, and onions at the beginning of 2019 and relatively high comprehensive price at the end of 2019.
While the comprehensive price for cucumbers, potatoes, and onions of the red-body markets tended to open higher, and finish lower.
\textbf{(2) For the OHLC data transformed by the second pseudo-PC scores (`Meat' factor)}, we see that the open price of Zhangjiajie is at the medium level. This indicates that at the beginning of 2019, the prices of beef, lamb, and pork in Zhangjiajie were not high.
However, its candlestick stretches to the highest and lowest among all the markets, suggesting that the `Meat' price in Zhangjiajie experienced great fluctuations in 2019, resulting in the prices appearing extremely high and extremely low.
On the other hand, the four-dimensional data of Jiangsu are very small, indicating that its price of `Meat' remained stable in 2019, which corresponds to the original lamb price of Jiangsu remained at $58 \; CNY/kg$ throughout the year.

It is important to mention that an eigenvector uniquely corresponds to one pseudo-PC and one transformed OHLC data.
However, for each eigenvalue of the standardized feature information matrix $\textrm{\textbf{Y}}$, there are two corresponding eigenvectors, which are mutually opposite vectors.
The difference in sign does not affect the orthogonal and normalized requirements in Theorem.\ref{Theorem_1}, they are both feasible solutions.
Paradoxically, the pseudo-PCs corresponding to these two eigenvectors are different, and the restored OHLC formed pseudo-principle component scores are also different.
In order to enhance the interpretability of the pseudo-PCA for OHLC data, it is recommended that the eigenvector is selected by the following rule.
If the eigenvector has the largest absolute loading values on the feature information variables, such as $\bm{Y}_1, \bm{Y}_2$ and $\bm{Y}_3$, then select the eigenvector that has positive loading values on $\bm{Y}_1, \bm{Y}_2$ and $\bm{Y}_3$.
This way, a relationship is established.
The original data size, feature information size, pseudo-PC scores, and OHLC formed pseudo-PC scores are all positively related to each other.
The advantage of this practice lies in the original data size and can be easily analyzed through the final comprehensive OHLC formed pseudo-principle component scores.
For instance, in this article, we see that the loading values of the `Vegetable' factor for cucumbers, potatoes, and onions are all positive;
the loading values of the `Meat' factor for beef, lamb, and pork are also all positive.

\vspace*{-0.8cm}
\section{Conclusions} \label{Sec 5}
\vspace*{-0.2cm}

As more OHLC data is collected, the demand for dimension reduction of the OHLC data emerges.
Inspired by the various PCA methods applied for interval data, this paper proposed a novel modus for characterizing features of OHLC data and built a unique $\mathbb{R}_n^{4}$ space containing the feature-based representations accordingly.
The new method of representing the OHLC data can eliminate the inherent constraints of OHLC data and transition between the original OHLC data and its feature-based representations freely.
In the unique $\mathbb{R}_n^{4}$ space, we defined the linear operations and inner product operation and deduced the pseudo-PCA method for OHLC data under this new algebraic framework.
The pseudo-PCA has high visibility and high interpretability, which can effectively identify important information and perform dimension reduction.
Therefore, reducing the dimensionality of an OHLC data set with $n$ observations and $p$ variables to a new OHLC data set with $n$ observations and $m$ components is achieved, and $m \leq p$.
The effectiveness, robustness and interpretability of the pseudo-PCA is examined by finite simulations and empirical experiment.

The OHLC formed pseudo-PCs derived from pseudo-PCA still hold the constraints of OHLC data, which maintains the original format of the raw data and provides significant convenience for the subsequent establishment of econometric or machine learning models.
Through the OHLC formed pseudo-PCs, one can observe the size and the fluctuation of certain original data variables during the observation period.
Furthermore, the OHLC formed pseudo-PCs can be presented in the format of the candlestick chart, which helps people obtain information easily.
Overall, this paper provides a novel feature-based representation method for OHLC data and enriches the existing literature on PCA methods.

For scalar data, the Singular Spectrum Analysis (SSA) is applied instead of PCA when performing dimension reduction of panel data.
Similarly, we may further propose the new pseudo-SSA approach to address the time-series OHLC data set, which can be an interesting research direction in the future.
Besides, exploring new feature extraction methods for more generalized interval-valued data is significant.
For instance, many types of data (i.e., daily temperatures, companies' profits, and stock returns) hold intervals and own values between the upper and lower boundaries (low value may be negative) while not belonging to OHLC data.

\vspace*{-0.8cm}
\section*{Declarations}
\vspace*{-0.5cm}

\section*{Funding}
\vspace*{-0.5cm}

This study was funded by the National Natural Science Foundation of China (grant numbers. $71420107025$, $11701023$).

\vspace*{-0.5cm}
\section*{Conflicts of interest}
\vspace*{-0.2cm}

Author Wenyang Huang declares that he has no conflict of interest.
Author Huiwen Wang declares that she has no conflict of interest.
Author Shanshan Wang declares that she has no conflict of interest.

\vspace*{-0.5cm}
\section*{Ethical approval}
\vspace*{-0.2cm}

This article does not contain any studies with human participants or animals performed by any of the authors.

\vspace*{-0.5cm}
\section*{Availability of data and material}
\vspace*{-0.2cm}

The spot data is downloaded form a financial server called Wind.
The data can be uploaded as required.

\vspace*{-0.5cm}
\section*{Code availability}
\vspace*{-0.2cm}

This paper applies custom R code and can be uploaded as required.

\vspace*{-0.5cm}
\nolinenumbers
\bibliography{KlinePCA}
\bibliographystyle{plainnat}
\clearpage
\appendixpage
\appendix
\renewcommand{\appendixname}{Appendix~\Alph{section}}

\section{Proof of Proposition.1}

\textbf{Proposition 1.} For any two OHLC data $\bm{x}$ and $\tilde{\bm{x}}$ with their corresponding feature-based representations $\bm{y}$ and $\tilde{\bm{y}}$, then it holds $\bm{x}=\tilde{\bm{x}}$ if and only if $\bm{y}=\tilde{\bm{y}}$ holds.
\vspace*{-0.5cm}

\begin{flushleft}
	$\bm{Proof:}$
\vspace*{-0.5cm}
\end{flushleft} Obviously, $\bm{x}=\tilde{\bm{x}}$ implies $\bm{y}=\tilde{\bm{y}}$ from Definition.2.
It suffices to prove that if $\bm{y}=\tilde{\bm{y}}$, then $\bm{x}=\tilde{\bm{x}}$.
In fact, since $\bm{y}=\left(y^{(1)}, y^{(2)}, y^{(3)}, y^{(4)}\right)^{\prime}$ and $\tilde{\bm{y}}=\left(\tilde{y}^{(1)}, \tilde{y}^{(2)}, \tilde{y}^{(3)}, \tilde{y}^{(4)}\right)^{\prime}$, $\bm{y}=\tilde{\bm{y}}$ implies that $y^{(j)}=\tilde{y}^{(j)}, \; j=1,2,3,4$.

First, $y^{(1)}=\text{ln}x^{(l)}$, $\tilde{y}^{(1)}=\text{ln}\tilde{x}^{(l)}$ $\Rightarrow$ $y^{(1)}=\tilde{y}^{(1)}$  implies $x^{(l)}=\tilde{x}^{(l)}$ (1).

Second, $y^{(2)}=\text{ln}(x^{(h)}-x^{(l)})$, then $y^{(2)}=\tilde{y}^{(2)}$ together with $x^{(l)}=\tilde{x}^{(l)}$ implies that $\tilde{y}^{(2)}=\text{ln}(\tilde{x}^{(h)}-\tilde{x}^{(l)})$, $x^{(h)}=\tilde{x}^{(h)}$ (2).

Third, $y^{(3)}=\tilde{y}^{(3)}$ and $y^{(4)}=\tilde{y}^{(4)}$ yield $\lambda^{(o)}=\tilde{\lambda}^{(o)}$ and $\lambda^{(c)}=\tilde{\lambda}^{(c)}$.

Thus, $x^{(o)}=\lambda^{(o)}x^{(h)}+(1-\lambda^{(o)})x^{(l)}=\tilde{\lambda}^{(o)}\tilde{x}^{(h)}+(1-\tilde{\lambda}^{(o)})\tilde{x}^{(l)}=\tilde{x}^{(o)}$ (3);
$x^{(c)}=\lambda^{(c)}x^{(h)}+(1-\lambda^{(c)})x^{(l)}=\tilde{\lambda}^{(c)}\tilde{x}^{(h)}+(1-\tilde{\lambda}^{(c)})\tilde{x}^{(l)}=\tilde{x}^{(c)}$ (4).

Therefore, $\bm{y}=\tilde{\bm{y}}$ $\Rightarrow$ $\bm{x}=\tilde{\bm{x}}$. As desired.
\begin{flushleft}
\vspace*{-0.5cm}
	$\bm{Proof \; up}$
\end{flushleft}

\clearpage

\section{Proof of Proposition.2}

\textbf{Proposition 2.} If the feature information variables $\bm{Y}_1, \bm{Y}_2,\ldots, \bm{Y}_p$ are standardized, the first $m$ pseudo-principle components $\bm{F}_h \in \mathbb{R}_n^4 \; ( 1 \leq h \leq m, m \leq p)$ derived from the pseudo-PCA satisfy the following properties:\\
(1)\; The sample mean of $\bm{F}_m$ is equal to $0$, that is, $\overline{\boldsymbol{F}}_{h}=\mathbf{0} \in \mathbb{R}^{4}$; \\
(2)\; The sample variance of $\bm{F}_m$ is equal to $\lambda_h$, namely, $\operatorname{Var}\left(\boldsymbol{F}_{h}\right)=\lambda_{h}$; \\
(3)\; For any $\bm{F}_h, \bm{F}_k \in \mathbb{R}_n^4$, if $h \neq k$, their sample covariance $Cov(\bm{F}_h, \bm{F}_k) = 0$; \\
(4)\; $\max \sum_{h=1}^{m} Var(\bm{F}_h)=p.$

\begin{flushleft}
\vspace*{-0.5cm}
	$\bm{Proof:}$
\vspace*{-0.5cm}
\end{flushleft}

In regard to conclusion (1), because it is assumed that $\bm{Y}_1, \bm{Y}_2,\ldots, \bm{Y}_p$ have been standardized, there is $\overline{\boldsymbol{Y}}_{j}=\mathbf{0}, \; j=1,2, \ldots, p$.
Since $\bm{F}_h$ is a linear combination of $\bm{Y}_1, \bm{Y}_2,\ldots, \bm{Y}_p$, it is easy to deduce:
\vspace{-2em}
\begin{align*}\label{Eq 31}
& \overline{\boldsymbol{F}}_{h}=E(\bm{F}_h)=E(u_{h 1} \otimes \boldsymbol{Y}_{1} \oplus u_{h 2} \otimes \boldsymbol{Y}_{2} \oplus \cdots \oplus u_{h p} \otimes \boldsymbol{Y}_{p})\\
& =u_{h 1} \times E(\boldsymbol{Y}_{1}) + u_{h 2} \times E(\boldsymbol{Y}_{2}) + \cdots + u_{h p} \times E(\boldsymbol{Y}_{p}) \\
& =u_{h 1} \times (0,0,0,0)' + u_{h 2} \times (0,0,0,0)' + \cdots + u_{h p} \times (0,0,0,0)' \\
& =(0,0,0,0)' + (0,0,0,0)' + \cdots + (0,0,0,0)'\\
& =(0,0,0,0)'\\
\end{align*}

\vspace{-4em}
As for conclusion (2), since $\bm{u}_h^\prime \bm{u}_h=1$, there is:
\begin{center}
	$\operatorname{Var}\left(\boldsymbol{F}_{h}\right)=\frac{1}{n}\left(\boldsymbol{F}_{h}, \boldsymbol{F}_{h}\right)_{\mathbb{R}_{n}^{4}}=\boldsymbol{u}_{h}^{\prime} \boldsymbol{W} \boldsymbol{u}_{h}=\boldsymbol{u}_{h}^{\prime} \lambda_{h} \boldsymbol{u}_{h}=\lambda_{h}$
\end{center}

The proof of conclusion (3) is similar.
For $h \neq k$, because of $\bm{u}_h$ and $\bm{u}_k$ are orthogonal, so:
\begin{center}
	$\operatorname{Cov}\left(\boldsymbol{F}_{h}, \boldsymbol{F}_{k}\right)=\frac{1}{n}\left(\boldsymbol{F}_{h}, \boldsymbol{F}_{k}\right)_{\mathbb{R}_{n}^{4}}=\boldsymbol{u}_{h}^{\prime} \boldsymbol{W} \boldsymbol{u}_{k}=\boldsymbol{u}_{h}^{\prime} \lambda_{k} \boldsymbol{u}_{k}=0$
\end{center}

When it comes to conclusion (4),
record matrix $\bm{U}=(\bm{u}_1,\bm{u}_2,\ldots,\bm{u}_p)$, which is an orthogonal matrix, and mark the diagonal matrix as $\boldsymbol{\Lambda}=\operatorname{diag}\left(\lambda_{1}, \lambda_{2}, \ldots, \lambda_{p}\right)$, because there is:
\begin{center}
	$\boldsymbol{U}^{\prime} \boldsymbol{W} \boldsymbol{U}=\boldsymbol{\Lambda}$
\end{center}
\vspace*{-0.5cm}
The correlation coefficient matrix $\bm{W}$ and $\boldsymbol{\Lambda}$ are two similar square matrices.
The traces of these two matrices are:
\vspace*{-0.5cm}
\begin{center}
	$\operatorname{tr}(\boldsymbol{W})=p, \quad \operatorname{tr}(\boldsymbol{\Lambda})=\sum_{h=1}^{p} \lambda_{h}$
\vspace*{-0.5cm}
\end{center}
Because the traces of similar square matrices are equal, and $Var(\bm{F}_h)=\lambda_h \; (h=1,2,\ldots,p)$, we can derive the conclusion (4): $\sum_{h=1}^{p} Var(\bm{F}_h)=p$.
\vspace*{-0.5cm}

\begin{flushleft}
	$\bm{Proof \; up}$
\end{flushleft}

\clearpage

\section{Data used in this paper}

\begin{table}[htbp]
	\caption {Raw OHLC data}
	\label{Raw OHLC data}
	\resizebox{\textwidth}{!}{
		\begin{tabular}{|c|c|c|c|c|c|c|}
			\hline
			\textbf{Markets}                                                                          & \textbf{Beef}               & \textbf{Lamb}             & \textbf{Pork}                 & \textbf{Cucumber}        & \textbf{Potato}         & \textbf{Onion}           \\ \hline
			Baotou Youyi market vegetable wholesale market                                            & (62, 72, 58, 72)'           & (57, 64, 52, 60)'         & (22, 56, 20, 44)'             & (5, 9, 1.4, 5)'          & (1.6, 3, 1.5, 2)'       & (1.4, 3, 1.4, 3)'        \\ \hline
			Beijing Dayanglu Agricultural and Sideline Products Market Co., Ltd                       & (53.5, 66.5, 52, 66)'       & (47, 54.5, 46.98, 54)'    & (15.8, 49.75, 11.5, 39)'      & (6.4, 9, 2, 6.6)'        & (2.6, 2.8, 1.8, 2.4)'   & (1.9, 3.3, 1.3, 3.2)'    \\ \hline
			Beijing Xinfadi Agricultural Products Co., Ltd                                            & (54.25, 67.5, 51.6, 65.75)' & (52.8, 56.9, 42.8, 54.8)' & (15.65, 50.25, 11.5, 40.5)'   & (4, 9, 1.2, 4.3)'        & (1.95, 3, 0.9, 2.3)'    & (1.3, 2.3, 0.95, 2.3)'   \\ \hline
			Changji Yuanfeng Agricultural and Sideline Products Trading Market Co., Ltd               & (60, 60.5, 56, 59)'         & (53, 61.5, 53, 60)'       & (19, 48.5, 18, 37)'           & (6, 12, 1.5, 7.5)'       & (1.8, 4, 1.5, 2)'       & (1.8, 3, 1, 2)'          \\ \hline
			Hefei Zhougudui Agricultural Products Wholesale Market Co., Ltd                           & (62, 74.5, 58, 74.5)'       & (61, 70, 54, 64)'         & (14.9, 43.14, 12.12, 36)'     & (5.5, 10, 2, 7)'         & (2.9, 3.7, 2.2, 2.7)'   & (1.8, 2.8, 1.2, 2.1)'    \\ \hline
			Jiangsu Lingjiatang Market Development Co., Ltd                                           & (70, 80, 58, 80)'           & (58, 58, 58, 58)'         & (22, 56.5, 19.8, 45)'         & (5.6, 6.6, 1.4, 4.5)'    & (1.8, 3.4, 1.6, 2.2)'   & (1.4, 3.6, 1.4, 3.6)'    \\ \hline
			Mawangdui Agricultural Products Co., Ltd                                                  & (66, 86, 66, 86)'           & (62, 73, 61, 73)'         & (18.8, 54, 16.9, 46)'         & (4, 7, 1.8, 3)'          & (2.6, 3.2, 1.6  2.6)'   & (1.8, 4.2, 1.3, 3.3)'    \\ \hline
			Inner Mongolia Dongwayao Agricultural and Sideline Products Wholesale Market Co., Ltd     & (62, 66.2, 56, 64)'         & (54, 60.5, 50, 56)'       & (24, 60, 18, 48)'             & (5, 10, 1.2, 5.6)'       & (2.2, 3.6, 1.2, 2.2)'   & (1.6, 3.1, 1.5,   2.8)'  \\ \hline
			Shanghai Jiangyang Agricultural Products Logistics Co., Ltd                               & (63, 80.4, 60, 79.2)'       & (56.1, 62.8, 54.4, 61.5)' & (19.39, 52.87, 13.14, 43.07)' & (7.2, 9.8, 2, 6.2)'      & (2.5, 3.4, 2, 2.4)'     & (2.2, 4.1, 1.6, 4.1)'    \\ \hline
			Tianjin Hanjiashu Haijixing Agricultural Products Logistics Co., Ltd                      & (57, 72, 55, 70)'           & (74, 82, 70, 78)'         & (17.3, 60, 17.3, 41)'         & (4.8, 10.2, 2, 5.7)'     & (2.3, 4, 1.4, 2.4)'     & (1.6, 4, 1.3, 3.5)'      \\ \hline
			Tianjin Jinyuanbao Binhai Agricultural Products Trading Market                            & (52, 58, 50, 58)'           & (48, 55, 48, 55)'         & (17.5, 61, 16, 43)'           & (4.3, 8.3, 1.2, 4.8)'    & (2.3, 3.6, 1.3, 2)'     & (1.3, 2.7, 1.2, 2.1)'    \\ \hline
			Xian Xinbeicheng Agricultural and Sideline Products Trading Market Management Co., Ltd    & (58.2, 68.2, 55, 68)'       & (58, 64.2, 51, 64)'       & (19.4, 52, 18, 42)'           & (5.2, 11, 2.2, 6)'       & (2.2, 3.6, 2, 2.6)'     & (1.8, 3.4, 1.6, 3)'      \\ \hline
			Xian Zhuque Agricultural Products Market Co., Ltd                                         & (60, 76.01, 56, 70)'        & (60, 76, 60, 62)'         & (17, 56, 17, 42)'             & (6, 11, 2, 9)'           & (2.6, 5, 2, 3)'         & (2.61, 3, 2, 3)'         \\ \hline
			Xinjiang Xibulvzhu Fruit and Vegetable Co., Ltd.                                          & (55, 63, 55, 63)'           & (52, 62, 52, 62)'         & (15.5, 47, 12.5, 39)'         & (5.5, 11, 1, 7.5)'       & (1.8, 3, 1.5, 2.2)'     & (1.3, 3, 1.2, 3)'        \\ \hline
			Xinyu Youzhi Agricultural Products Wholesale Market                                       & (63.7, 80.1, 63.7, 80)'     & (63.9, 68, 60, 66)'       & (23.4, 58.5, 22.4, 56)'       & (4.8, 10, 3.4, 6)'       & (4.2, 5, 3.8, 4.1)'     & (4.5, 4.8, 3.6, 4.2)'    \\ \hline
			Xinyang Yunong Agricultural Products Sales Co., Ltd                                       & (45, 48, 42, 47)'           & (55, 57, 52, 56)'         & (19, 44, 16.5, 39)'           & (4.3, 6.1, 3.1, 5.5)'    & (2, 2.6, 2, 2.3)'       & (2.1, 2.2, 1.3, 1.6)'    \\ \hline
			Suqian Nancaishi Agricultural and Sideline Products Wholesale Market Management Co., Ltd  & (60.95, 75, 60, 74)'        & (56.05, 78, 56, 77)'      & (17.68, 54, 17.02, 44)'       & (4.31, 7.95, 1.02, 7.5)' & (2.7, 3.85, 2, 3.7)'    & (1.58, 2.82, 1.02, 2.8)' \\ \hline
			Yinchuan Beihuan Vegetables and Fruits Comprehensive Wholesale Market Management Co., Ltd & (54.6, 69, 51, 68)'         & (57.4, 66, 50, 64)'       & (19.2, 56, 16, 46)'           & (3.7, 9.04, 0.76, 5.8)'  & (1.52, 2.16, 1.2, 1.5)' & (1.2, 1.6, 0.86, 0.92)'  \\ \hline
			Zhangjiajie Yongding District Market Management Service Center                            & (56, 98, 54, 80)'           & (56, 94, 54, 74)'         & (27, 70, 23, 46)'             & (6, 10, 4, 4.2)'         & (5, 8, 4, 4)'           & (7, 10, 4, 4)'           \\ \hline
			Changzhi Zifang Agricultural Products Comprehensive Trading Market Co., Ltd               & (64, 74, 60, 72)'           & (72, 74, 66, 72)'         & (18, 64, 16, 50)'             & (4.8, 12, 1.6, 5.4)'     & (1.8, 3.2, 1.4, 1.8)'   & (1.4, 3.6, 1, 3.6)'      \\ \hline
	\end{tabular}}
\end{table}

\begin{table}[htbp]
	\caption {Feature-based representations}
	\label{Feature information data}
	\resizebox{\textwidth}{!}{
		\begin{tabular}{|c|c|c|c|c|c|c|}
			\hline
			\textbf{Markets}                                                                          & \textbf{Beef}                & \textbf{Lamb}                 & \textbf{Pork}                & \textbf{Cucumber}            & \textbf{Potato}               & \textbf{Onion}               \\ \hline
			Baotou Youyi market vegetable wholesale market                                            & (0.03, 0, 0.63, 1.2)'        & (-0.02, 0.13, 0.89, -0.71)'   & (0.27, -0.06, 0.36, -0.34)'  & (-0.27, 0.05, 0.22, -0.29)'  & (-0.13, -0.13, -1.67, -0.02)' & (0, -0.05, -2.58, 1.21)'     \\ \hline
			Beijing Dayanglu Agricultural and Sideline Products Market Co., Ltd                       & (-0.05, 0.02, -0.2, 0.03)'   & (-0.07, -0.11, -1.92, 0.27)'  & (-0.47, 0.02, 1.39, -0.01)'  & (0.25, -0.07, 1.14, 0.81)'   & (0.04, -0.52, 2.22, 1.05)'    & (-0.05, 0.07, 0.27, 0.8)'    \\ \hline
			Beijing Xinfadi Agricultural Products Co., Ltd                                            & (-0.05, 0.08, 0.17, -0.81)'  & (-0.12, 0.21, 1.51, -0.18)'   & (-0.47, 0.04, 1.32, 0.2)'    & (-0.49, 0.09, -0.48, -0.74)' & (-0.63, 0.2, 0.88, 1.32)'     & (-0.24, -0.17, 0.13, 1.42)'  \\ \hline
			Changji Yuanfeng Agricultural and Sideline Products Trading Market Co., Ltd               & (0, -0.77, 2.65, -1.75)'     & (-0.01, -0.04, -1.67, -0.28)' & (0.13, -0.28, -0.38, -0.59)' & (-0.17, 0.52, -0.05, 0.29)'  & (-0.13, 0.36, -1.05, -0.69)'  & (-0.21, 0.07, 0.55, -1.07)'  \\ \hline
			Hefei Zhougudui Agricultural Products Wholesale Market Co., Ltd                           & (0.03, 0.11, 0.48, 1.4)'     & (0, 0.27, 0.93, -0.8)'        & (-0.4, -0.26, 1.05, 0.36)'   & (0.25, 0.12, 0, 0.61)'       & (0.24, -0.13, 0.75, -0.02)'   & (-0.1, -0.07, 0.49, -0.91)'  \\ \hline
			Jiangsu Lingjiatang Market Development Co., Ltd                                           & (0.03, 0.3, 1.37, 2.27)'     & (0.03, -2.08, 1.09, -1.03)'   & (0.26, -0.04, 0.47, -0.21)'  & (-0.27, -0.5, 2.46, 0.43)'   & (-0.07, 0.05, -1.13, -0.02)'  & (-0.04, 0.15, -1.21, 2.6)'   \\ \hline
			Mawangdui Agricultural Products Co., Ltd                                                  & (0.11, 0.24, -2.34, 1.78)'   & (0.06, 0.13, -0.15, 4.49)'    & (0.05, -0.02, 0.24, 0.47)'   & (0.1, -0.5, -0.08, -1.89)'   & (-0.07, -0.07, 1.38, 1.15)'   & (-0.05, 0.3, -0.18, -0.56)'  \\ \hline
			Inner Mongolia Dongwayao Agricultural and Sideline Products Wholesale Market Co., Ltd     & (0, -0.22, 1.49, -1.34)'     & (-0.04, 0.06, 0.82, -0.91)'   & (0.13, 0.15, 1.76, -0.04)'   & (-0.49, 0.26, -0.03, -0.13)' & (-0.35, 0.32, 0.56, 0.33)'    & (0.04, -0.07, -0.9, -0.14)'  \\ \hline
			Shanghai Jiangyang Agricultural Products Logistics Co., Ltd                               & (0.05, 0.24, 0.07, -0.35)'   & (0, -0.05, 0.37, -0.2)'       & (-0.29, 0.07, 1.91, 0.23)'   & (0.25, 0.09, 1.38, 0.09)'    & (0.14, -0.2, 0.31, -0.23)'    & (0.09, 0.23, 0.05, 0.97)'    \\ \hline
			Tianjin Hanjiashu Haijixing Agricultural Products Logistics Co., Ltd                      & (-0.01, 0.12, -0.1, -0.86)'  & (0.13, 0.13, 0.71, -0.71)'    & (0.08, 0.17, -5.64, -0.97)'  & (0.25, 0.16, -0.59, -0.42)'  & (-0.2, 0.4, 0.27, 0.2)'       & (-0.05, 0.26, -0.51, -0.13)' \\ \hline
			Tianjin Jinyuanbao Binhai Agricultural Products Trading Market                            & (-0.07, -0.38, 0.5, 0.75)'   & (-0.06, -0.13, -1.91, 1.23)'  & (-0.03, 0.24, -0.36, -0.72)' & (-0.49, -0.05, 0, -0.09)'    & (-0.27, 0.28, 0.63, -0.15)'   & (-0.1, -0.12, -0.86, -0.81)' \\ \hline
			Xian Xinbeicheng Agricultural and Sideline Products Trading Market Management Co., Ltd    & (-0.01, -0.05, 0.49, 0.6)'   & (-0.03, 0.18, 1.12, 1.04)'    & (0.13, -0.14, -0.06, -0.09)' & (0.39, 0.26, -0.59, -0.53)'  & (0.14, -0.07, -1, 0.16)'      & (0.09, 0, -0.51, -0.27)'     \\ \hline
			Xian Zhuque Agricultural Products Market Co., Ltd                                         & (0, 0.23, 0.32, -1.64)'      & (0.05, 0.28, -1.89, -2.02)'   & (0.05, 0.05, -4.12, -0.48)'  & (0.25, 0.3, 0.04, 1.69)'     & (0.14, 0.54, -0.46, -0.02)'   & (0.23, -0.37, 1.08, 2.13)'   \\ \hline
			Xinjiang Xibulvzhu Fruit and Vegetable Co., Ltd.                                          & (-0.01, -0.38, -3.48, 2.35)' & (-0.02, 0.04, -2.37, 1.83)'   & (-0.36, -0.12, 1.01, 0.34)'  & (-0.76, 0.45, 0.08, 0.77)'   & (-0.13, -0.13, -0.46, 0.52)'  & (-0.1, 0.02, -1.01, 1.01)'   \\ \hline
			Xinyu Youzhi Agricultural Products Wholesale Market                                       & (0.09, 0.1, -2.93, 1.22)'    & (0.05, -0.07, 1.03, -0.5)'    & (0.43, -0.06, -0.61, 2.23)'  & (1.02, -0.16, -1.54, -0.76)' & (0.77, -0.35, 0.21, -0.41)'   & (0.6, -0.26, 1.51, -1.07)'   \\ \hline
			Xinyang Yunong Agricultural Products Sales Co., Ltd                                       & (-0.19, -0.58, 1.25, -1.13)' & (-0.02, -0.31, 1.26, -0.36)'  & (0.02, -0.42, 1.07, 0.75)'   & (0.89, -1.31, -0.22, 1.89)'  & (0.12, -0.94, -1.54, 0.8)'    & (-0.05, -0.44, 2.13, -1.51)' \\ \hline
			Suqian Nancaishi Agricultural and Sideline Products Wholesale Market Management Co., Ltd  & (0.05, 0.04, -0.56, -0.44)'  & (0.02, 0.43, -2, 0.47)'       & (0.06, -0.03, -1.22, 0.07)'  & (-0.73, -0.09, 0.22, 3.75)'  & (0.14, 0.07, 0.4, 3)'         & (-0.2, 0, 0.31, 1.77)'       \\ \hline
			Yinchuan Beihuan Vegetables and Fruits Comprehensive Wholesale Market Management Co., Ltd & (-0.06, 0.16, 0.32, -0.31)'  & (-0.04, 0.27, 0.98, -0.08)'   & (-0.03, 0.08, 0.89, 0.21)'   & (-1.16, 0.17, -0.5, 0.51)'   & (-0.35, -0.56, 0.21, -0.11)'  & (-0.31, -0.56, 0.71, -2.6)'  \\ \hline
			Zhangjiajie Yongding District Market Management Service Center                            & (-0.02, 0.76, -0.79, -1.96)' & (0, 0.73, -0.42, -1.06)'      & (0.46, 0.3, 0.98, -1.33)'    & (1.26, -0.3, -0.64, -5.04)'  & (0.81, 0.82, -0.18, -6.31)'   & (0.66, 0.77, 0.82, -4.36)'   \\ \hline
			Changzhi Zifang Agricultural Products Comprehensive Trading Market Co., Ltd               & (0.05, -0.01, 0.64, -1.01)'  & (0.1, -0.07, 1.61, -0.5)'     & (-0.03, 0.33, -0.05, -0.08)' & (-0.07, 0.51, -0.81, -0.94)' & (-0.2, 0.05, -0.33, -0.56)'   & (-0.21, 0.24, -0.28, 1.5)'   \\ \hline
	\end{tabular}}
\end{table}

\clearpage

\section{Abbreviations}

\begin{table}[htbp]
	\caption {Markets' full name-abbreviated comparison table}
	\label{Market full name-abbreviated comparison table}
	\resizebox{\textwidth}{!}{
		\begin{tabular}{|c|c|}
			\hline
			\textbf{Markets}                                                                          & \textbf{Abbreviations} \\ \hline
			Baotou Youyi market vegetable wholesale market                                            & Baotou                \\ \hline
			Beijing Dayanglu Agricultural and Sideline Products Market Co., Ltd                       & Beijing1              \\ \hline
			Beijing Xinfadi Agricultural Products Co., Ltd                                            & Beijing2              \\ \hline
			Changji Yuanfeng Agricultural and Sideline Products Trading Market Co., Ltd               & Changji               \\ \hline
			Hefei Zhougudui Agricultural Products Wholesale Market Co., Ltd                           & Hefei                 \\ \hline
			Jiangsu Lingjiatang Market Development Co., Ltd                                           & Jiangsu               \\ \hline
			Mawangdui Agricultural Products Co., Ltd                                                  & Mawangdui             \\ \hline
			Inner Mongolia Dongwayao Agricultural and Sideline Products Wholesale Market Co., Ltd     & Inner Mongolia        \\ \hline
			Shanghai Jiangyang Agricultural Products Logistics Co., Ltd                               & Shanghai              \\ \hline
			Tianjin Hanjiashu Haijixing Agricultural Products Logistics Co., Ltd                      & Tianjin1              \\ \hline
			Tianjin Jinyuanbao Binhai Agricultural Products Trading Market                            & Tianjin2              \\ \hline
			Xian Xinbeicheng Agricultural and Sideline Products Trading Market Management Co., Ltd    & Xian1                 \\ \hline
			Xian Zhuque Agricultural Products Market Co., Ltd                                         & Xian2                 \\ \hline
			Xinjiang Xibulvzhu Fruit and Vegetable Co., Ltd.                                          & Xinjiang              \\ \hline
			Xinyu Youzhi Agricultural Products Wholesale Market                                       & Xinyu                 \\ \hline
			Xinyang Yunong Agricultural Products Sales Co., Ltd                                       & Xinyang               \\ \hline
			Suqian Nancaishi Agricultural and Sideline Products Wholesale Market Management Co., Ltd  & Suqian                \\ \hline
			Yinchuan Beihuan Vegetables and Fruits Comprehensive Wholesale Market Management Co., Ltd & Yinchuan              \\ \hline
			Zhangjiajie Yongding District Market Management Service Center                            & Zhangjiajie           \\ \hline
			Changzhi Zifang Agricultural Products Comprehensive Trading Market Co., Ltd               & Changzhi              \\ \hline
	\end{tabular}}
\end{table}

\end{document}